\documentclass[preprint]{elsarticle}

\usepackage{natbib, graphicx, hyperref}
\usepackage{amsmath}
\usepackage{amssymb}
\usepackage[dvipsnames]{xcolor}
\usepackage{textcomp}

\newcommand{\norm}[1]{\left\lVert#1\right\rVert}
\journal{Journal of Computational Physics}

\begin{document}
\begin{frontmatter}
\title{Comparison of Split-Step and Hamiltonian Integration
Methods for Simulation of the Nonlinear Schr\"{o}dinger Type Equations}

    \author[UNM]{Anastassiya A.~Semenova\corref{cor1}}
    \ead{asemenov@math.unm.edu}
    \cortext[cor1]{Corresponding author}

    \address[UNM]{Department of Mathematics \& Statistics, The University of New Mexico,
    MSC01 1115, 1 University of New Mexico, Albuquerque, New Mexico, 87131-0001, USA}

    \author[UW,UB]{Sergey A.~Dyachenko}
    \ead{sergd@uw.edu}

    \address[Landau]{L.\,D.~Landau Institute for Theoretical Physics,
    2 Kosygin Str., Moscow, 119334, Russian Federation}
    \address[UW]{Department of Applied Mathematics, University of Washington, Lewis Hall 201, Box 353925, Seattle, Washington 98195-3925, USA}
    \address[UB]{Department of Mathematics, SUNY Buffalo, 244 Mathematics Building, Buffalo, NY 14260-6284, USA}

    \author[UNM,Landau]{Alexander O.~Korotkevich}
    \ead{alexkor@math.unm.edu}

    \author[UNM,Landau,VShE]{Pavel M.~Lushnikov}
    \ead{plushnik@math.unm.edu}
    \address[VShE]{NRU Higher School of Economics, Myasnitskaya 20, Moscow, 101000, Russian Federation}

\begin{abstract}
We provide a systematic comparison of two numerical methods to solve the widely used nonlinear Schr\"{o}dinger equation (NLSE).
The first one is the standard second order split-step (SS2) method based on operator splitting approach. The second one is the Hamiltonian
integration method (HIM), originally proposed in the paper by Dyachenko et al in 1992 (Physica D, vol. 57, pp. 97-160). Extension of the HIM to a widely used generalization of NLSE is developed. HIM allows the exact conservation of the Hamiltonian
and wave action at
the cost of requiring iterative solution for the implicit scheme.
The numerical error for HIM is smaller than the SS2 solution for the same time step for
almost all simulations we consider. Conversely,
one can take orders of magnitude larger time steps in HIM, compared with SS2, still ensuring numerical stability.

\end{abstract}

\end{frontmatter}


\section{Introduction}
A nonlinear Schr\"{o}dinger equation
(NLSE) is one of the most generic  nonlinear partial differential equation in numerous branches of mathematical and theoretical physics~\cite{sulem1999nonlinear}. NLSE naturally appears
if one considers envelope dynamics of a quasi--monochromatic nonlinear wave~\cite{ZakharovEtAl2009} in a system, where the first nonlinear correction to dispersion relation is proportional to intensity. In quantum mechanics  a version of NLSE is called a Gross--Pitaevskii equation \cite{pitajevskij2003bose} which describes a Bose-Einstein condensate with a short-range interactions of particles.

A typical NLSE application is the dynamics of optical pulses in an optical fiber. The time evolution of the envelope of an optical pulse
in a fiber is well approximated by NLSE, including the description of very long, transoceanic optical communication
lines, see e.g.~\cite{agrawal2007nonlinear,PLOpticsLetters2001}.
The Langmuir waves in plasmas are described by NLSE as well, see e.g.~\cite{zakharov1972collapse,SilantyevLushnikovRosePartIPhysPlasm2017}. Dynamics of quasi-monochromatic oceanic waves (which is typical e.g. for ocean swell) is reduced to NLSE or its modifications \cite{dysthe1979note}. For example, the analysis of NLSE offers a possible explanation to the mystery of appearance of the rogue waves~\cite{dyachenko2005modulation}.
All these and numerous other applications of NLSE and its modification require efficient numerical simulation.

Many techniques can be used in simulation of NLSE: the Crank-Nicholson scheme, the hopscotch method, the Ablowitz--Ladik scheme,
the pseudo--spectral split-step method, the Hamiltonian preserving method, and many others
(see~\cite{taha1984analytical},
~\cite{greig20ll},~\cite{richtmyer1994difference}).
One of the most popular methods of integration of NLSE, called split-step, was proposed by F.~Tappert~\cite{Tappert}, and its performance was studied in~\cite{taha1984analytical}. The split-step method can be considered as a version of
the Strang's operator splitting approach~\cite{strang1968construction} combined with pseudo-spectral method. The split-step method can be constructed
to any order of accuracy, in this work we consider the second order symmetrized split-step (SS2) method as the most popular one.
A recent study of stability of the split-step method can be found in the work~\cite{Lakoba2012} and references therein.

In 1992 a novel method for simulation of NLSE has been proposed in the paper~\cite{dyachenko1992optical}. It has been successful to study turbulence in
two--dimensional NLSE, however it passed largely unnoticed by a wide audience. Perhaps, that is the reason why it was not mentioned in the recent papers
such as~\cite{chen2002symplectic},~\cite{gong2017conservative}, which describe somewhat similar numerical methods. This numerical method, which we call the Hamiltonian
integration method (HIM), conserves the numerical Hamiltonian and the optical power (also called number of particles or wave action) exactly (in exact arithmetic), and it is based on discrete Hamilton's equations.
In finite precision arithmetics, the error in conservation of Hamiltonian is due to round-off errors inherent to specific finite precision floating point representation.

By using the discrete Hamilton's equations in other systems one may derive Hamiltonian-preserving numerical schemes. As an example, we refer the reader to
the recent work~\cite{korotkevich2016numerical} on numerical simulations of nonlinear water waves. One can trace similarities with the symplectic methods~\cite{Yoshida1990},
while HIM is a completely self-contained {\it ad hoc} method which can be derived for other Hamiltonian systems having canonical symplectic structure.
For example, we have done it for Maija, McLaughlin, and Tabak (MMT) model~\cite{initialMMT} which is a widely used generalization of NLSE.

We compare the two numerical methods by performing a set of simulations with various initial conditions. In these experiments we observe that in some
scenarios HIM method can outperforms SS2 when very high accuracy is not essential. The SS2 method requires a stringent condition on time step for stability,
whereas HIM is an implicit method and as such allows the time step to be a hundred times larger. Our observations illustrate that HIM method can be the method
of choice for efficient simulations of interaction of solitons, where a tight balance between nonlinearity and dispersion occurs.

The paper is organized as follows: we describe the mathematical problem in section~\ref{s1}; the description of numerical methods is given in section~\ref{s2};
the section~\ref{PhysicalPara} discusses the relation between the dimensionless NLSE and the physical units relevant to optical fibers communications;
implementation of HIM for MMT is given in the section~\ref{MMTscheme};
the section~\ref{s3} describes the set of simulations and discusses obtained results; and in section~\ref{s4} we summarize our observations and discuss the
applicability of both methods. The derivation of HIM method is placed in~\ref{appA} and the convergence conditions are discussed in~\ref{appB}.
The results of simulations of head on collision of solitons
and collision with pursuing soliton are presented in~\ref{appC}.

\section{\label{s1} Problem Formulation}
Let us consider NLSE in its simplest form (rescaling of coordinate, time, and amplitude can bring NLSE into this form without loss of generality):
\begin{equation}
i\Phi_t+\Phi_{xx}+\gamma|\Phi|^2\Phi=0,\label{eq:nlse}
\end{equation}
where $\Phi(x,t)$ is a complex function, $\gamma = \pm 1$ denotes focusing and defocusing NLSE respectively, and subscript denotes partial derivative with respect to $x$ and $t$. The latter
equation is solved on an interval $x\in[-L,L]$ subject to periodic boundary conditions, and $t\in[0,T]$.
For the sake of simplicity, we consider NLSE in one spatial dimension, although both methods are applicable to any dimensions (for example, HIM was originally formulated for $2\text{D}$ problem~\cite{dyachenko1992optical}).
\subsection{Constants of Motion}

The Hamiltonian, $\mathcal{H}$, and the number of particles, $\mathcal{N}$, given by:
\begin{align}
    \mathcal{H} = \int \Big( |\Phi_x|^2- \frac{\gamma}{2} |\Phi|^4 \Big) dx\quad\mbox{and}\quad
    \mathcal{N} = \int |\Phi|^2\,dx,\label{eq:Hamiltonian}
\end{align}
are conserved quantities for~\eqref{eq:nlse}. Here and further we integrate over one spatial period $[-L,L]$ and drop the integration limits for brevity.
The NLSE is an integrable system \cite{shabat1972exact}, and it has infinitely many nontrivial integrals of motion, that may be used to track accuracy
of numerical simulation. We consider first two nontrivial integrals of motion, that are given by \cite{shabat1972exact}, \cite{novikov1984theory}:
\begin{align}
    \mathcal{C}_4 &= \int \left[ \Phi {\bar\Phi}_{xxx}+\frac{3\gamma}{2}\Phi{\bar{\Phi}}_x{|\Phi|}^2 \right]\, dx, \\
    \mathcal{C}_5 &= \int \left[ |\Phi_{xx}|^2 + \frac{\gamma^2}{2}|\Phi|^6 - \frac{\gamma}{2}\left( |\Phi|^2_x \right)^2 - 3\gamma|\Phi|^2|\Phi_x|^2 \right]\, dx.
\end{align}
We denote them $\mathcal{C}_4$ and $\mathcal{C}_5$ because the first three are so called trivial integrals of motion:
the number of particles $\mathcal{N}$ and the Hamiltonian $\mathcal{H}$~\eqref{eq:Hamiltonian}, and the momentum.

\subsection{Exact Solutions of NLSE}
The equation~\eqref{eq:nlse}, has soliton solutions \cite{shabat1972exact},
and when NLSE is considered on infinite spatial interval, it may be solved
by means of the inverse scattering transform (IST). Some solutions of NLSE that decay at $x\to\pm\infty$, such as $N$-soliton
solutions may be used on a periodic interval when the magnitude of $|\Phi|$ is close enough to zero at the endpoints $x=\pm L$.

The one-soliton solution (here and further we use $\gamma = 1$) is given by the formula:
\begin{align}
    \Phi = \dfrac{\sqrt{2\lambda}e^{i\left(\frac{1}{2}vx+\left(\lambda-\frac{1}{4}v^2\right)t+\Phi_0\right)}}{\cosh{\left[\sqrt{\lambda}\left(x-vt-x_0\right)\right]}},\label{1soliton}
\end{align}
where $x_0$, and $v$ are the constants that determine initial position and the propagation speed of the soliton, and the
constants $\lambda$ and $\Phi_0$ determine the soliton amplitude and the initial phase respectively.

Another exact solution of~\eqref{eq:nlse} on infinite line is the two-soliton solution which can be obtained by dressing method \cite{zakharov1974scheme}, given by the formula:
\begin{align}
    \Phi &= \dfrac{ \left[1 + \dfrac{e^{\eta_2 + \bar\eta_2}(p_1-p_2)^2}{2(p_1 + \bar p_2)^2(p_2 + \bar p_2)^2}\right]e^{\eta_1} +
                    \left[1 + \dfrac{e^{\eta_1 + \bar\eta_1}(p_1-p_2)^2}{2(\bar p_1 + p_2)^2(p_1 + \bar p_1)^2}\right]e^{\eta_2}}
               { D },\label{2soliton}
\end{align}
where $D$ is the following expression:
\begin{align}
    D &= 1 + \dfrac{e^{\eta_1 + \bar\eta_1}}{2(p_1 + \bar p_1)^2} +
             \dfrac{e^{\eta_2 + \bar\eta_2}}{2(p_2 + \bar p_2)^2} +
         \dfrac{e^{\eta_1 + \bar\eta_2}}{2(p_1 + \bar p_2)^2} +
         \dfrac{e^{\bar\eta_1 + \eta_2}}{2(\bar p_1 + p_2)^2} + \nonumber \\
      &\qquad\qquad\qquad\qquad\qquad\qquad + \dfrac{e^{\eta_1 + \bar\eta_1 + \eta_2 + \bar\eta_2}|p_1-p_2|^4}{4(p_1+\bar p_1)^2(p_2 + \bar p_2)^2|p_1 + \bar p_2|^4}
\end{align}
and $\eta_1$,$\eta_2$ are determined by the expression:
\begin{align}
\eta_{1,2} = p_{1,2}\,x + ip_{1,2}^2\,t + a_{1,2},
\end{align}
here $p_{1,2}$ and $a_{1,2}$ are complex constants. The width and the propagation speed of solitons are defined by
the real and the imaginary parts of $p_{1,2}$ respectively. The initial positions of each soliton are defined by $a_{1,2}$.

\subsection{Numerical Solution on Periodic Interval}
It is natural to use Fourier series to approximate $\Phi(x,t)$ on the periodic interval $x\in[-L,L]$ using a pseudo spectral approach by the means of the discrete Fourier transform (DFT) that is computed using the fast Fourier transform library FFTW~\cite{fftw3}. In
physical space we use a uniform grid,
\begin{align}
    x_j = \frac{2L}{N}j - L\quad\mbox{where $j = 0,\ldots N-1$}
\end{align}
to discretize the interval $[-L,L]$. We introduce a grid function, $\Phi_j^{n} = \Phi(x_j, n \Delta t)$, where $\Delta t$ is an
elementary time step.

\section{\label{s2} Description of Numerical Methods}
\subsection{The SS2 Method}
In the SS2 method, the linear and nonlinear terms of (\ref{eq:nlse}) are treated separately in a style of Strang splitting \cite{strang1968construction}.

Let $\hat{L} = i{\partial}^2/{\partial x}^2$ represent the operator for the linear term and $\hat{N} = i\gamma{|\Phi|}^2$ represent the operator for the nonlinear term of (\ref{eq:nlse}), then $\Phi_t(x,t)=(\hat{L}+\hat{N})\Phi(x,t)$.
This equation has the formal solution $\Phi(x,t+\Delta t)=e^{(\hat{L}+\hat{N})\Delta t}\Phi(x,t)$ on a time step $\Delta t$. In the SS2 method \cite{taha1984analytical} we approximate the exponential term by the product of separate exponents:
\begin{align}\label{expTerm}
    e^{(\hat{L}+\hat{N})\Delta t} = e^{\hat{L} \frac{\Delta t}{2}} e^{\hat{N}\Delta t}e^{\hat{L} \frac{\Delta t}{2}}+\frac{\Delta t^3}{12}\{[\hat{L},[\hat{N},\hat{L}]]+\frac{1}{2}[\hat{N},[\hat{N},\hat{L}]]\}+\dots ,
\end{align}
that is accurate up to third order in time, and here $[\hat A,\hat B] = \hat A \hat B - \hat B \hat A$ defines the
commutator of operators $\hat A$ and $\hat B$.
This is a special case of application of Campbell-Baker-Hausdorff formula \cite{CBH}.
By doing this, the evolution of the linear part and  nonlinear part on the step $\Delta t$ can be carried out separately. In the context of NLSE this is particularly attractive because both evolutions can be carried out analytically.
Note that the linear PDE $i\Phi_t= -\Phi_{xx}$, can be solved exactly in the Fourier domain:
\begin{align}\label{eq:ss2L}
\Phi_k(t+\Delta t) = e^{-ik^2\Delta t}\Phi_k(t),
\end{align}
where $\Phi_k(t)$ denotes the Fourier coefficient, corresponding to wavenumber $k$, of $\Phi(x,t)$.
The nonlinear part of~\eqref{eq:nlse} given by $i\Phi_t =-\gamma |\Phi|^2 \Phi$ is an ODE, and can be solved exactly:
\begin{align}\label{eq:ss2NL}
    \Phi(x,t+\Delta t) = e^{i\gamma|\Phi|^2\Delta t}\Phi(x,t).
\end{align}
Equations~\eqref{eq:ss2L} and~\eqref{eq:ss2NL} give us explicit expressions for $e^{\hat{L}}$ and $e^{\hat{N}}$ correspondingly. The only complexity is that these two exact solutions are given in Fourier and coordinate spaces which requires switching between them in order to represent $e^{\hat{L} \frac{\Delta t}{2}} e^{\hat{N}\Delta t}e^{\hat{L} \frac{\Delta t}{2}}$ in~\eqref{expTerm} consecutively.

In a similar manner one may construct higher order split step methods, by alternating linear and nonlinear steps. The SS2 method is stable if the condition,
\begin{equation}\label{eq:ss2}
    \Delta t \leq \frac{{\Delta x}^2}{\pi}
\end{equation}
described in \cite{weideman1986split} is satisfied. However, one can violate this condition when the highest Fourier coefficients are small enough.

One can note that both steps (linear and nonlinear one) in SS2 methods are performing only rotation of phase, so conservation of number of particle $\mathcal{N}$
is an intrinsic property of the method.

\subsection{Hamiltonian Integration Method}
The main feature of the HIM method (introduced in~\cite{dyachenko1992optical}) is its exact conservation of
the Hamiltonian, $\mathcal{H}$, and number of particles, $\mathcal{N}$. This is achieved by requiring that
the difference in $\mathcal{H}$ (and $\mathcal{N}$) on subsequent time steps vanishes, the
details of derivation of HIM are given in the~\ref{appA}. HIM is an implicit scheme:
\begin{equation}\label{eq:eq2}
    i\frac{\Phi^{n+1}_j-\Phi^n_j}{\Delta t}=-\frac{\left[\Phi^{n+1}_j + \Phi^n_j\right]_{xx}}{2}-
\frac{(\Phi^{n+1}_j+\Phi^n_j)(|\Phi^{n+1}_j|^2+|\Phi^n_j|^2)}{4}.
\end{equation}
that is solved by means of fixed point iterations on every time step
Equation~\eqref{eq:eq2} implicitly defines the solution at the subsequent time steps.
In the Fourier space the formula~\eqref{eq:eq2} becomes the following:
\begin{equation}\label{eq:neq}
\hat \Phi^{n+1}_k- \hat \Phi^n_k = -\frac{ik^2\Delta t}{2}(\hat \Phi^{n+1}_k+\hat \Phi^n_k)
    +\frac{i\Delta t}{4} \hat{F}\left[(\Phi^{n+1}+\Phi^n)(|\Phi^{n+1}|^2+|\Phi^n|^2)\right],
\end{equation}
where $\hat \Phi^n_k = \hat F\left[\Phi^n \right]$ is the $k$-th Fourier coefficient of the grid function $\Phi^n_j$.
Following the work~\cite{korotkevich2016numerical}, the linear  part of the equation~\eqref{eq:neq} can be resolved for $\hat \Phi^{n+1}_k$ which yields:
\begin{equation}\label{eq:eq3}
\hat \Phi^{n+1}_k=\frac{1-i\frac{k^2 \Delta t}{2}}{1+i\frac{k^2 \Delta t}{2}}\hat \Phi_k^n+i\frac{\Delta t}{4(1+i\frac{k^2 \Delta t}{2})}\hat{F}
    \left[(\Phi^{n+1}+\Phi^n)(|\Phi^{n+1}|^2+|\Phi^n|^2)\right].
\end{equation}
The equation~\eqref{eq:eq3} can be solved by fixed point iterations:
\begin{align}\label{eq:eq4}
    \hat \Phi^{n+1,s+1}_k&=\frac{1-i\frac{k^2 \Delta t}{2}}{1+i\frac{k^2 \Delta t}{2}}\hat \Phi_k^{n}
    + \frac{i\Delta t}{4(1+i\frac{k^2 \Delta t}{2})}\hat{F}\left[(\Phi^{n+1,s}+\Phi^{n})(|\Phi^{n+1,s}|^2+|\Phi^{n}|^2)\right],
\end{align}
where $s$ denotes the iteration number and $\hat \Phi^{n+1,0}_k = \hat \Phi^n_k$. We iterate~\eqref{eq:eq4} until the residual condition is satisfied:
\begin{align}
    \norm{\hat \Phi_k^{n+1,s+1} -\hat \Phi_k^{n+1, s}}_{2} = \sqrt{\sum_{k} \left|\hat \Phi_k^{n+1,s+1} -\hat \Phi_k^{n+1, s} \right|^2 }  \leq \varepsilon \, , \label{Residual}
\end{align}
where $\norm{\cdot}_2$ denotes the $l_2$ norm on $[-L,L]$, and $\varepsilon$ is the tolerance for fixed point iterations. The initial
values $\Phi^{n+1,0}$ are computed by using one step of Forward Euler. Following \cite{dyachenko1992optical}, the fixed point iterations of HIM converge
for
\begin{align}
    \Delta t < \dfrac{2}{\sqrt 3\max\limits_{j}({|\Phi_j^n|}^2)}.\label{NLSEconvcond}
\end{align}
Derivation of this condition is given in~\ref{appB}.

For the time step that satisfies the above condition, the fixed point iterations typically converge in $4$ to $6$ steps
with the tolerance $\varepsilon \leq 10^{-11}$.
\section{Physical Units Relevant to Optical Fiber\label{PhysicalPara}}
Before the investigation of the performance of the two methods on a long time scale, we would
like to estimate the characteristic time of simulation that corresponds to the dynamics of a
pulse in a physically realistic fiber. In order to do so we consider a trans--Atlantic fiber
described in the reference paper~\cite{PLOpticsLetters2001} subject to:
\begin{equation}\label{physicnlse}
    iA_z - \frac{1}{2} \beta_2 A_{\tau \tau} + \sigma_1 |A|^2 A = 0 \, .
\end{equation}
We use the values for $\beta_2 = -20\,\mbox{ps}^2\,\mbox{km}^{-1}$, the group velocity dispersion (GVD), and $\sigma_1  = 1.3\times10^{-3}\,\mbox{km}^{-1}\mbox{mW}^{-1}$, the strength of nonlinearity for a fiber, provided therein.

The dimensionless NLSE given by~\eqref{eq:nlse} must be rewritten in the original dimensional
units. We transform the dimensionless NLSE to dimensional units as follows:
\begin{equation}\label{ztaua}
    z = l t, \, \, \tau =  \frac{x}{\omega_0},\mbox{ and } A = A_0\Phi
\end{equation}
The derivatives with respect to $t$ and $x$ are given by:
\begin{equation}
    \partial_t = l \partial_z \mbox{ and } \partial_x = \frac{1}{\omega_0}\partial_{\tau}.
\end{equation}
The resulting equation transforms into:
\begin{equation}\label{nlsenew}
i A_z+\frac{1}{\omega_0^2l}A_{\tau\tau}+\frac{|A|^2A}{A_0^2l} = 0.
\end{equation}
Comparison of two equations~\eqref{physicnlse} and~\eqref{nlsenew} reveals that:
\begin{align}
    \beta_2 \left[\frac{ps^2}{km}\right] = \frac{-2}{\omega_0^2l}\,\, , \label{parameters1}\\
    \sigma_1 \left[\frac{1}{km \; mW}\right] = \frac{1}{A_0^2l}\,\, ,\label{parameters2}
\end{align}
where $A_0 = 1\,\mbox{mW}^{1/2}$. By using the parameters $\beta_2$ and $\sigma_1$
from the reference paper~\cite{PLOpticsLetters2001}, we find that $l\approx 769\,\mbox{km}$, $\omega_0^2 = 1.3\times 10^{-4}\,\mbox{ps}^{-2}$ from the equations~\eqref{parameters1}--\eqref{parameters2}.
We find that it is necessary to simulate the fiber until the dimensionless time $t_{max} \approx 13$ in order to mimic a $10^4$ km fiber. The nonlinear time is then given by $t_{NL} = \frac{\pi}{|\Phi|^2} = \frac{\pi}{2|\lambda|}$ which in physical units corresponds to $z_{NL} = t_{NL}l$.

\section{Hamiltonian Integration Method for MMT Model\label{MMTscheme}}
In 1997 a new model of one-dimensional dispersive wave turbulence was introduced by Majda, McLaughlin, and Tabak~\cite{initialMMT}. The MMT equation is given by:
\begin{equation}
i\psi_t=|\partial_x|^{\alpha}\psi+\gamma|\partial_x|^{-\beta/4}\left(||\partial_x|^{-\beta/4}\psi|^2|\partial_x|^{-\beta/4}\psi\right),\label{eq:MMT}
\end{equation}
and it can be considered as a generalization of a NLSE. Here $\alpha>0$ and $\beta$ are real parameters. This model describes a Hamiltonian system with $\mathcal{H}$ given by:
\begin{equation}
    \mathcal{H}_{MMT} = \int \left( ||\partial_x|^{\alpha/2}\psi|^2+ \frac{\gamma}{2} ||\partial_x|^{-\beta/4}\psi|^4 \right) dx.\label{eq:MMTHamiltonian}
\end{equation}
The MMT conserves number of particle, or wave action $\mathcal{N}$ similar to NLSE. One may note that for $\alpha=2$ and $\beta=0$ MMT is almost identical to
NLSE (a derivative $\partial_x$ is replaced by a nonlocal operator $|\partial_x|$ in the kinetic energy~\eqref{eq:MMTHamiltonian}, which results in the opposite sign in
front of the linear term of~\ref{eq:MMT}). The MMT model is widely used (see e.g. \cite{ZVD2001}, \cite{Lee3237}, \cite{RUMPF20131260})  for investigation of the wave turbulence theory~\cite{ZLF1992} for 2D hydrodynamics with 1D free surface. MMT equation is an
example of a system for which the same approach as in~\ref{appA} can be used, resulting in the following numerical scheme for HIM:
\begin{align}\label{eq:MMTeq2}
    &i\frac{\psi^{n+1}_j-\psi^n_j}{\Delta t}=\frac{|\partial_x|^{\alpha}\psi^{n+1}_j + |\partial_x|^{\alpha}\psi^n_j}{2}+\\ \nonumber
    &+\gamma|\partial_x|^{-\beta/4}\left(\frac{|\partial_x|^{-\beta/4}\psi^{n+1}_j+|\partial_x|^{-\beta/4}\psi^n_j}{2}\frac{||\partial_x|^{-\beta/4}\psi^{n+1}_j|^2+||\partial_x|^{-\beta/4}\psi^n_j|^2}{2}\right).
\end{align}
Similarly to NLSE both Hamiltonian $\mathcal{H}_{MMT}$ and number of particles (wave action) $\mathcal{N}$ are conserved exactly.
Solving for $\psi^{n+1}_j$ in the linear part of~\eqref{eq:MMTeq2} and applying the same approach as in~\ref{appB} one can get the following convergence condition:
\begin{equation}
    \Delta t < \dfrac{2}{|\gamma|\sqrt{3}k_{max}^{\beta/2}\max\limits_{j}({||\partial_x|^{-\beta/4}\Phi_j^n|^2})},
\end{equation}
where $k_{max}$ is the maximum of the absolute value of wave number. As one can see, if $\beta=0$ it coincides with convergence condition~\eqref{NLSEconvcond} of HIM for NLSE.

\section{\label{s3} Numerical Methods Performance}
In exact arithmetic, HIM conserves Hamiltonian $\mathcal{H}$ and optical power $\mathcal{N}$ up to any precision governed by the tolerance
threshold chosen for fixed point iterations, and SS2 conserves the number of particles exactly by the construction of the method.
However, in double precision the error in conservation of Hamiltonian and the optical power is due to
round-off errors inherent to floating point arithmetic. The round-off error accumulates in time and causes the
optical power for SS2 and HIM, and Hamiltonian for HIM to change.

\subsection{Stationary One-Soliton Solution}
\begin{figure}
\includegraphics[width=0.495\textwidth]{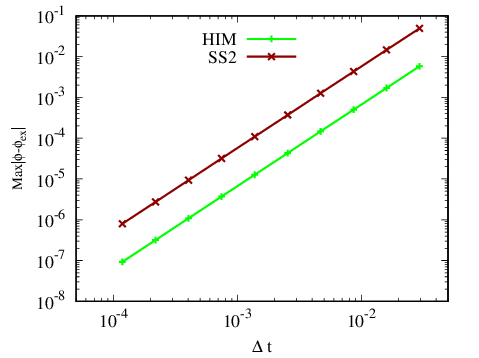}
\includegraphics[width=0.495\textwidth]{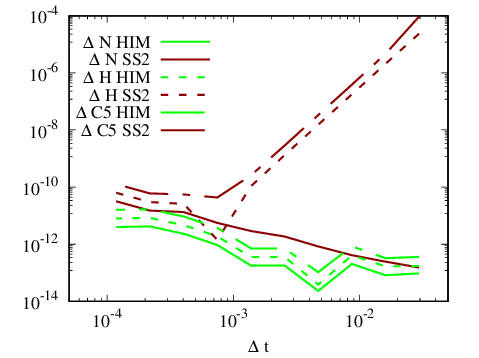}
    \caption{(Stationary one-soliton solution on a fully resolved grid) (Left) Convergence rate of numerical methods, HIM (green) and SS2 (red). Both methods have        second order convergence, but $\mathcal{L}_{\infty}$ error in solution is about one order smaller for HIM compared to SS2 for the same time steps.
    (Right) Error in conserved
    quantities: number of particles $\mathcal{N}$ (solid), Hamiltonian  $\mathcal{H}$ (dotted), and  $\mathcal{C}_5$(dash-dotted) for various time steps.
    When time step is larger than the stability condition of SS2, errors in $\mathcal{H}$ and $\mathcal{C}_5$ start to grow.
    For HIM, the error is dominated by accumulation of round-off errors and is smaller by several orders of magnitude compared with SS2.
    }
\label{fig:convergence}
\end{figure}

In this simulation we check the convergence rate of HIM and SS2 by running a sequence of simulations with various time steps. As the
initial condition we consider a one-soliton solution~\eqref{1soliton} with the following parameters:
\begin{align}
    \lambda = 2, \quad\mbox{and}\quad \Phi_0 = x_0 = v = 0.
\end{align}
We run the simulation on a fully resolved (highest harmonics are of round-off level) uniform grid of $N=2048$ grid points, and $L = 25\pi$.
The tolerance for HIM iterations
is set to $\varepsilon = 10^{-15}$ and simulation time is $T = 5$. The convergence of both methods is demonstrated in Figure~\ref{fig:convergence}.
We omit the $\mathcal{C}_4$ in the Figure~\ref{fig:convergence} because this quantity is identically zero for a stationary one-soliton
solution. The error in the integrals of motion for SS2 method is dominated by accumulation of round-off errors for small $\Delta t$, and
by the order of method for large $\Delta t$ as shown in the Figure~\ref{fig:convergence}. The critical value of $\Delta t$ for
which the transition occurs is close to the stability condition of SS2 method.

\subsection{Moving One-Soliton Solution}
\begin{figure}
\includegraphics[width=0.32\textwidth]{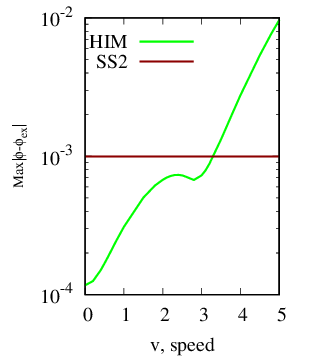}
\includegraphics[width=0.32\textwidth]{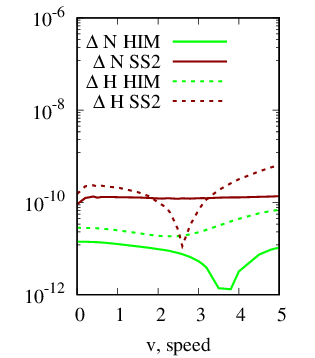}
\includegraphics[width=0.32\textwidth]{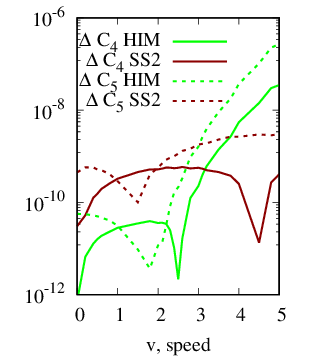}
    \caption{(Moving one-soliton solution on a fully resolved grid) (Left) The maximum absolute error of the solution at time $T=100$ as a function of propagation speed of the soliton.
    The SS2 method (red) has no dependence of the error on travel speed of the soliton because it naturally captures the dispersion
    relation of NLSE, while HIM (green) has dispersion relation accurate up to $\Delta t^3$. (Center) The error in integral
    quantities, $\mathcal{N}$~(solid), and $\mathcal{H}$~(dotted) is about seven orders of magnitude smaller than the error in
    the solution. (Right) The error in integral quantities, $\mathcal{C}_4$~(solid),
    and $\mathcal{C}_5$~(dotted) is about seven orders of magnitude smaller than the error in the solution.
    For travel speed $v \leq 3$ HIM and SS2 give comparable accuracy in $\mathcal{C}_4$ and $\mathcal{C}_5$, but HIM behaves worse
    as soon as $v$ is larger than $3$.  }
\label{fig:moving_error}
\end{figure}

In these simulations we investigate how the traveling speed $v$ of the one-soliton solution~\eqref{1soliton} affects the accuracy of both
numerical methods. It is known that dispersion of waves by SS2 method is identical to the dispersion of NLSE, while from~\eqref{eq:eq3} it
follows that the dispersion of HIM is only accurate up to third order in $k^2 \Delta t$. We expect that for sufficiently large time step
the travel speed of soliton will deviate from its true value. We show the results of the simulations with various travel speeds in
Figure~\ref{fig:moving_error}. The initial data for these simulations is given by~\eqref{1soliton} with parameters:
\begin{align}
    \lambda = 2, \quad\mbox{and}\quad \Phi_0 = x_0 = 0,\quad\mbox{and}\quad v \in[0,5].
\end{align}
The computational box size is $L = {25}\pi$ and the number of grid points is $N = 2048$. The tolerance for HIM iterations is $\varepsilon =
10^{-14}$ and the simulation time is $T = 100$. Time step for both methods is set to be $\Delta t = \frac{0.5 \Delta x^2}{\pi}$.

It should be noted, that soliton velocity is given in dimensionless units. In the left panel of Figure~\ref{fig:moving_error}, we
observe that the error in the solution has no dependence on travel speed of the soliton for SS2 method. For HIM, the error in the
solution depends on travel speed which is due to inexact dispersion relation of HIM method:
\begin{align}
    \omega_{HIM}(k) = \frac{i}{\Delta t} \ln\frac{1 - i \frac{k^2\Delta t}{2}}{1 + i \frac{k^2\Delta t}{2}} = k^2\left( 1 - \frac{k^4\Delta t^2}{12} + \ldots \right),
\end{align}
where $\omega_{HIM}(k)$ is the angular frequency of the $k$--th Fourier harmonic.

In the center panel, we look at the absolute error in integral quantities, $\mathcal{N}$ and $\mathcal{H}$.
It is about seven orders of magnitude smaller than the error in the solution.
On the right panel, we consider the absolute error in integral quantities, $\mathcal{C}_4$
and $\mathcal{C}_5$. Similarly to $\mathcal{N}$ and $\mathcal{H}$, it is about seven orders
of magnitude smaller than the error in the solution.
We notice that for travel speed $v \leq 3$ HIM and SS2 give comparable accuracy in
$\mathcal{C}_4$ and $\mathcal{C}_5$, but the error in HIM becomes
larger as soon as $v$ gets larger than $3$.
We see dips in the error of integral quantities as a function of speed. The
magnitude of the dips  is about one order, and it has no correlation to the error in
solution, which is significantly larger.
Note that the error in the solution does not always correlate with the error in integral quantities.

\subsection{Stationary Two-Soliton Solution}
\begin{figure}
\includegraphics[width=2.4in]{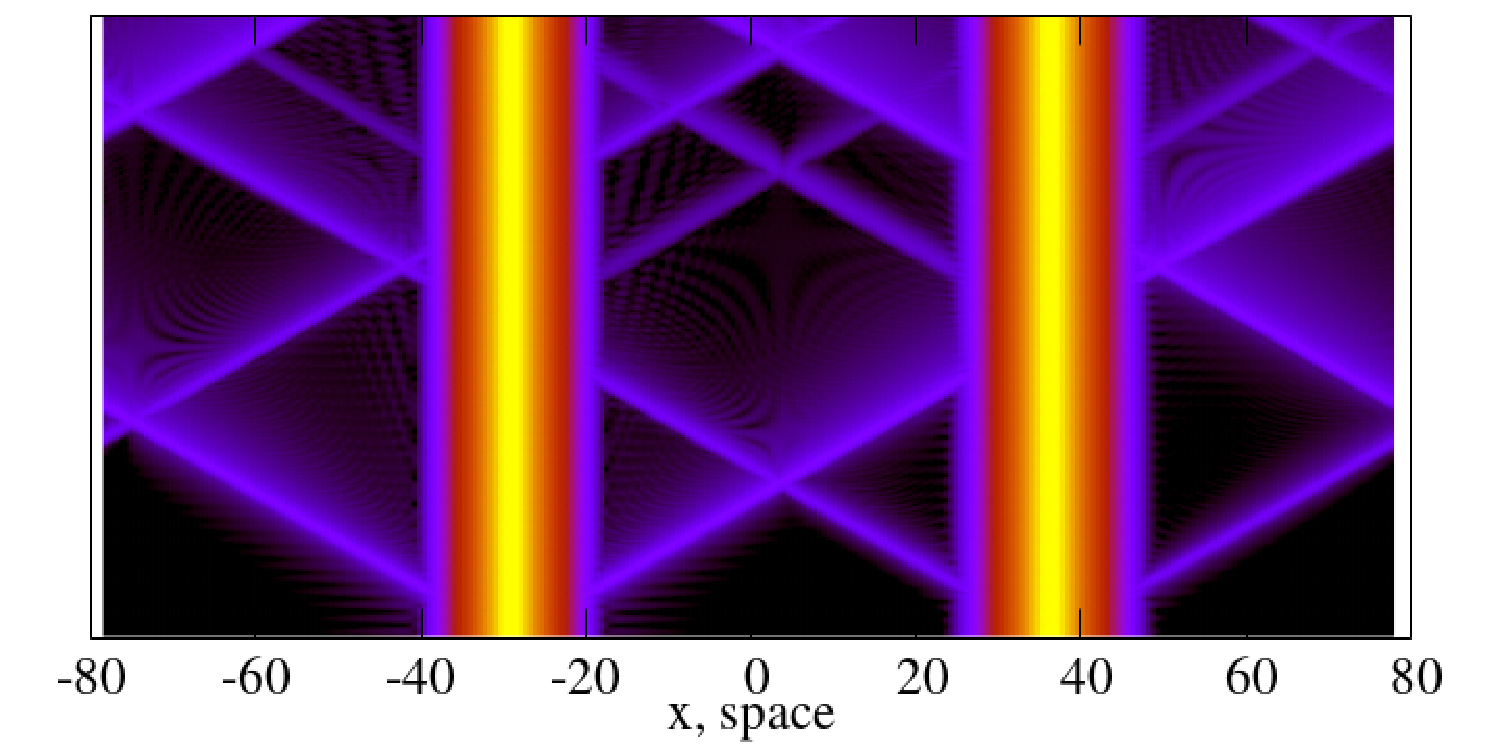}
\includegraphics[width=2.4in]{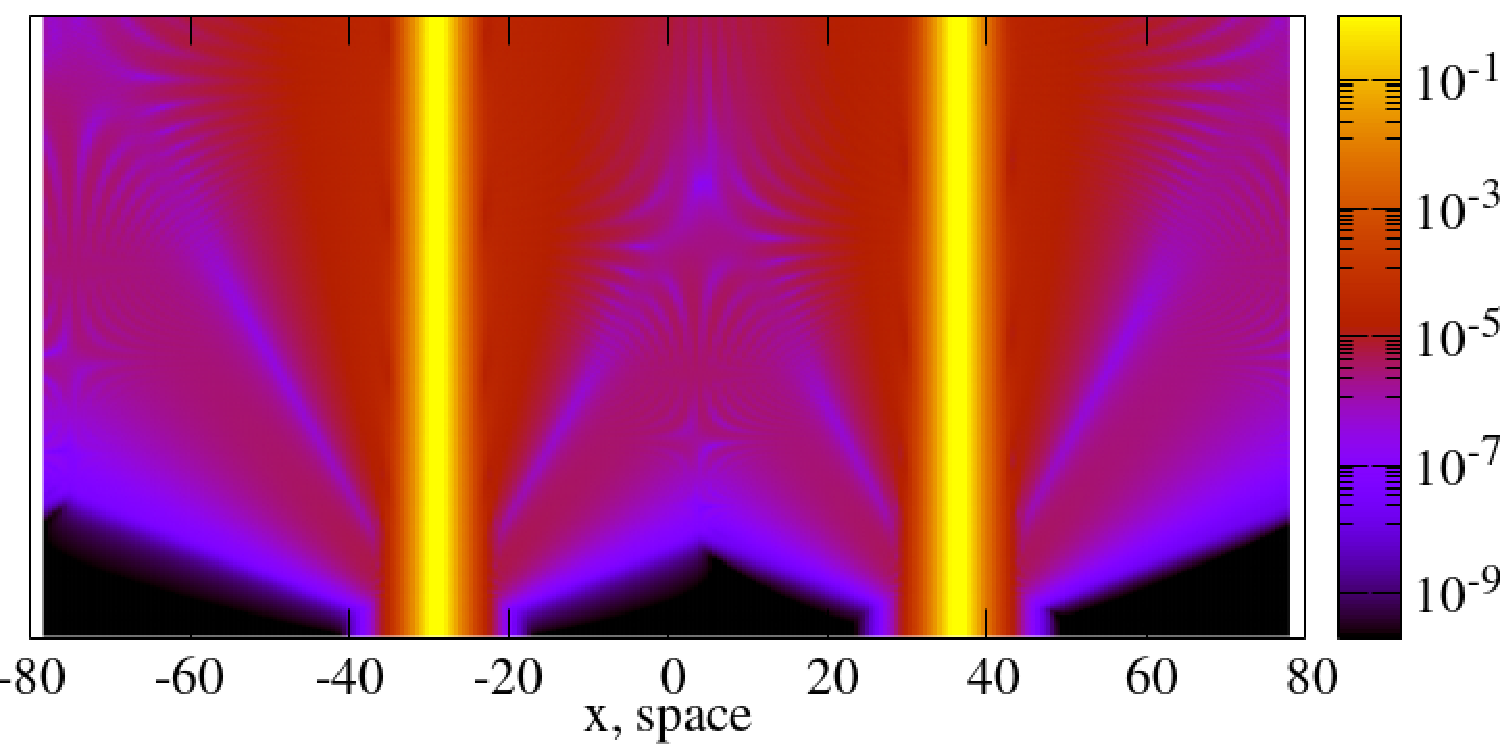}
\caption{(Stationary Two-Soliton Solution) Solution of NLSE with the initial data~\eqref{ic1} with HIM method (left), and SS2 method (right).
    The SS2 method radiates waves continuously over the course of the simulation, while the HIM emits
    localized small amplitude perturbations that travel in the computational box and are reflected and
    transmitted through the stationary solitons. At the time $T = 5$, the background radiation around the
    stationary solitons emitted in SS2 is several orders of magnitude
    larger than for HIM.}
\label{fig1:xtplane}
\end{figure}

\begin{figure}
\includegraphics[width=0.495\linewidth]{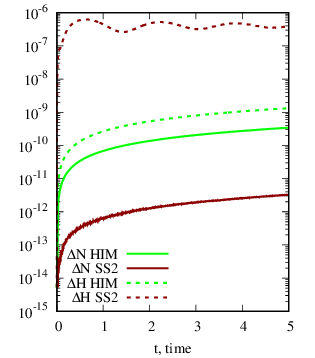}
\includegraphics[width=0.495\linewidth]{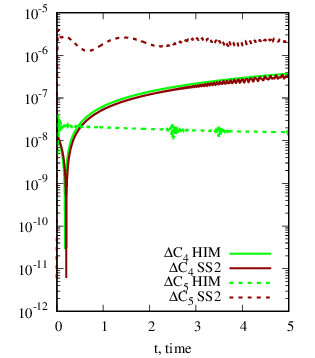}
\caption{(Stationary two-soliton solution on underresolved grid)
    Conserved integrals in a simulation with initial data~\eqref{ic1}.
    (Left) The number of particles (solid) and the Hamiltonian (dotted) computed
    via SS2 (red) and HIM (green). (Right) The integrals $\mathcal{C}_4$(solid)
    and $\mathcal{C}_5$ (dotted) via SS2 (red) and HIM (green).}
    \label{fig1:2SolConstantIntegrals}
\end{figure}

In this simulation we demonstrate the difference between SS2 and HIM when the initial data is a two-soliton solution with the following set of
parameters:
\begin{align}
p_1 = 2.0\,\,\mbox{and}\,\,p_2 = 1.9 \nonumber \\
a_1 = 60+i = -\bar a_2. \label{ic1}
\end{align}
The simulation time is $T = 5$, the solution is underresolved on a  grid with $N = 1024$ points. The computation box is $x \in [-L,L]$ where $L = 25\pi$.
The time step is $\Delta t = \frac{0.5 \Delta x ^2}{\pi}$.
It is typical to have solution not resolved to round--off error in long and/or multichannel simulations of light pulses propagating in optical fibers.
A smaller number of Fourier harmonics implies faster computations. For this experiment, the smallest amplitudes were of the order $10^{-8}$.
We present the results of the simulation in Figures~\ref{fig1:xtplane} -~\ref{fig1:2SolConstantIntegrals}.

In the course of simulation we observe that the error in $\mathcal{H}$ and $\mathcal{C}_5$ is one to two orders of
magnitude smaller in HIM than in SS2. The number of particles is better conserved by SS2 and the error is two orders
of magnitude smaller.

\subsection{Interaction of Two-Solitons}
In this section, we study the dynamics of the two-soliton solution~\eqref{2soliton}. We present parameters of
simulations in sections~\ref{CollisionStatSoliton} and show the results for the case of collision of a stationary
and a moving soliton. 

The results are similar to the other two cases: the head--on collision of solitons, and the collision with a
pursuing soliton. The latter ones are discussed in ~\ref{appC}.

In all three cases, we use periodic box with $L = 25\pi$ and $N = 4096$ grid points for fully resolved simulations and $N = 1024$ for unresolved simulations.
The time step is $\Delta t = \frac{0.8\Delta x ^2}{\pi} < \frac{\Delta x^2}{\pi}$ to satisfy the stability condition~\eqref{eq:ss2} in all three simulations.
The HIM iterations tolerance is $\epsilon = 10^{-12}$.

\subsubsection{Collision with Stationary Soliton}\label{CollisionStatSoliton}
\begin{figure}
\includegraphics[width=2.4in]{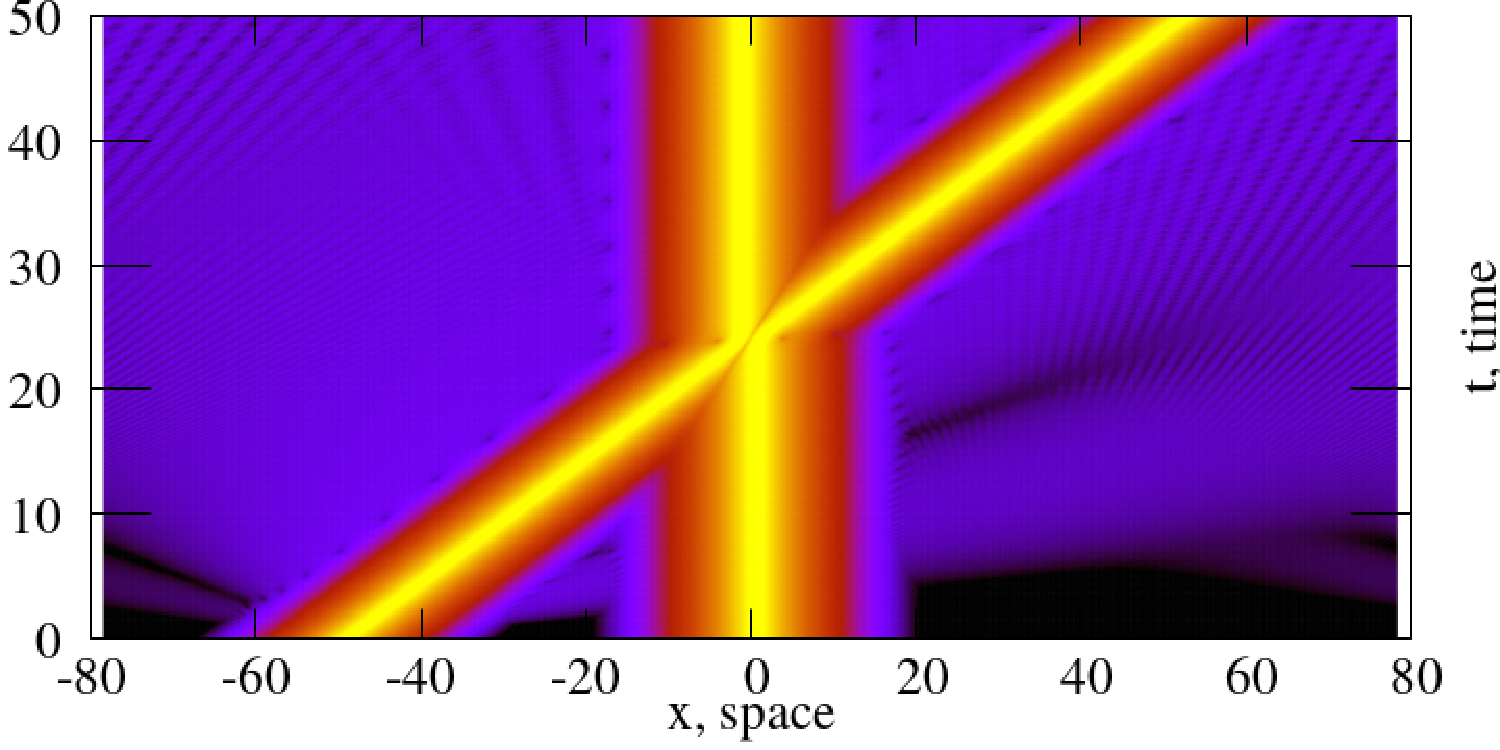}
\includegraphics[width=2.4in]{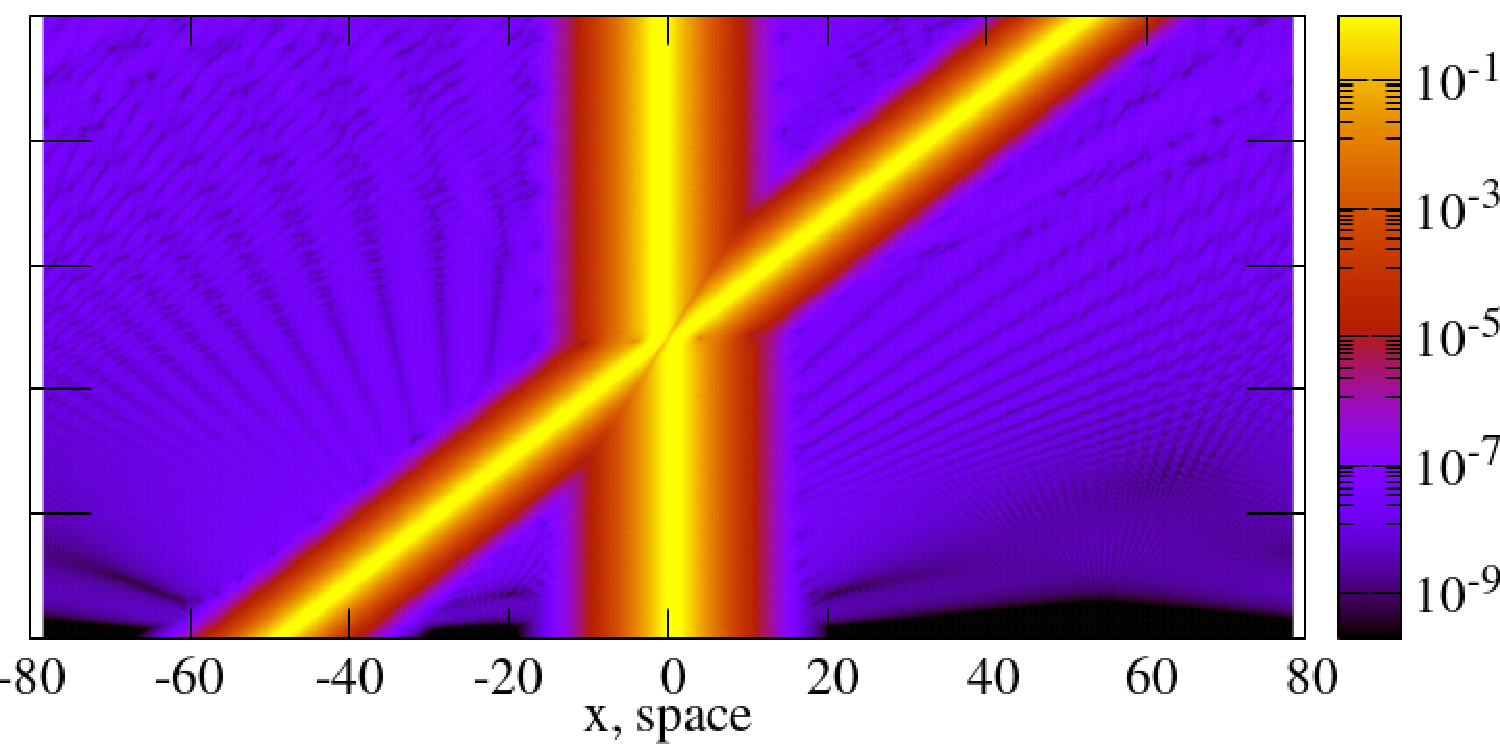}
\includegraphics[width=2.4in]{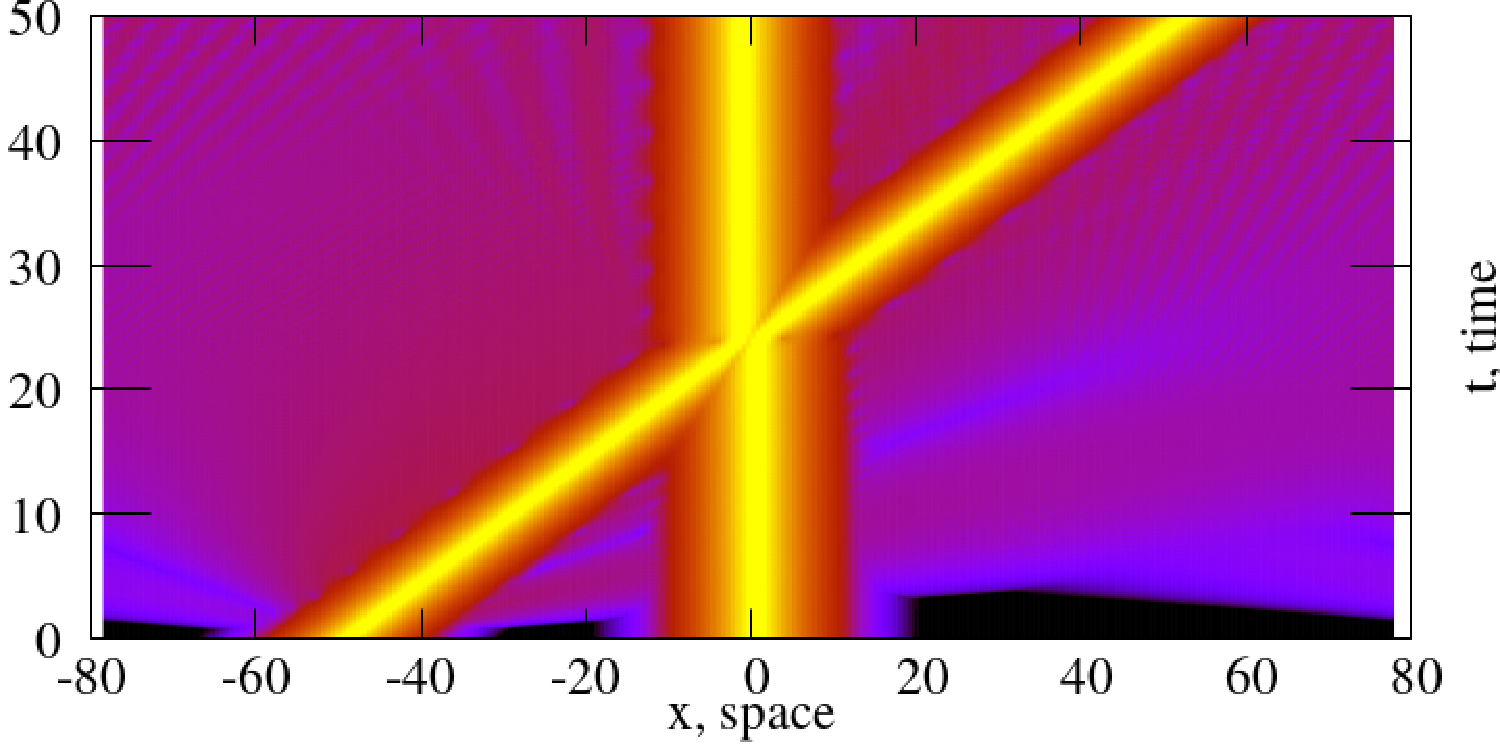}
\includegraphics[width=2.4in]{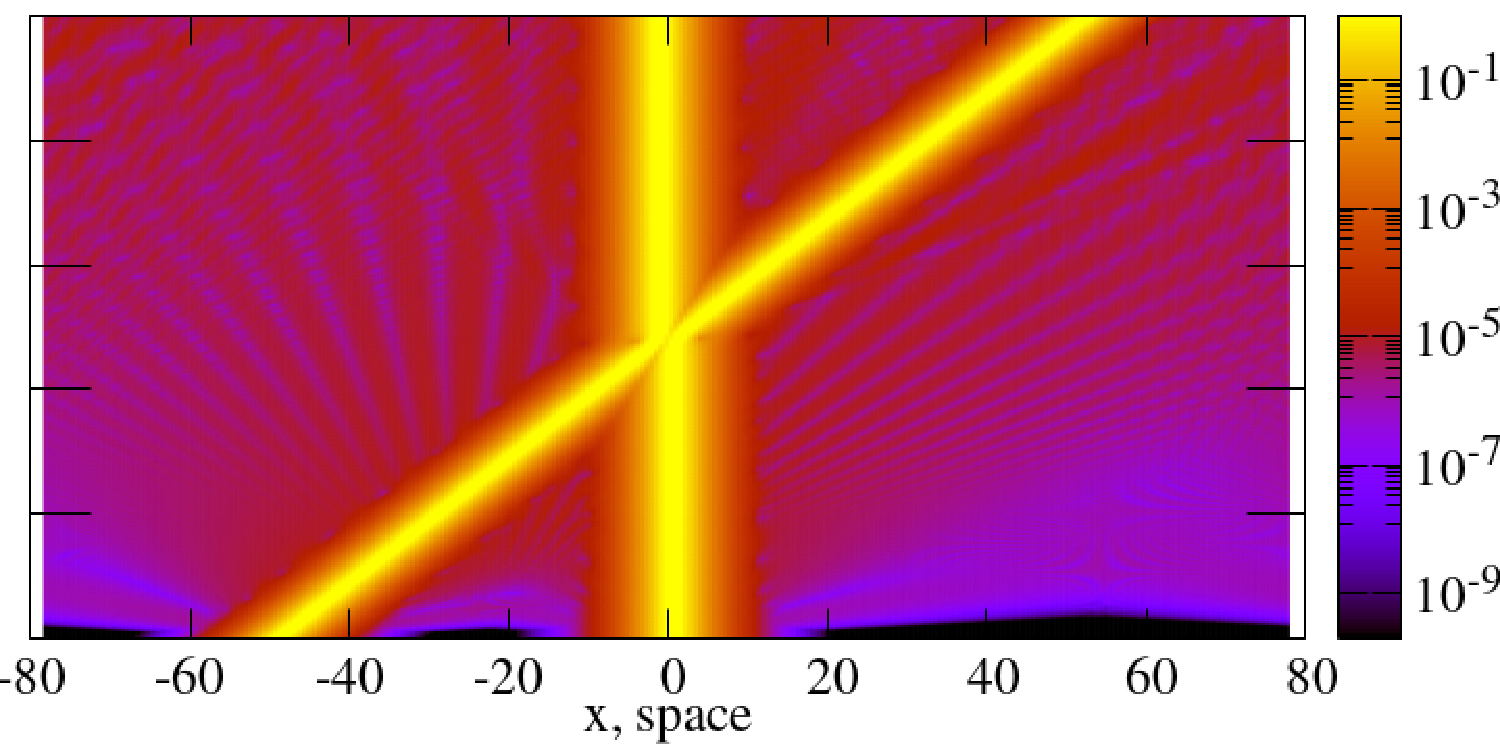}
    \caption{(Collision with stationary soliton) (Top) Numerical solution for HIM (left) and SS2 (right) methods on a fully resolved grid $N = 4096$.
(Bottom) Numerical solution for  HIM (left) and SS2 (right) methods on an underresolved grid with $N = 1024$.}
\label{fig:xtnabegaet}
\end{figure}

\begin{figure}
\includegraphics[width=0.495\linewidth]{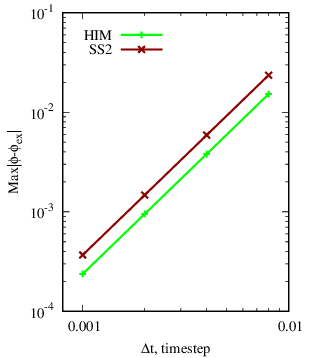}
\includegraphics[width=0.495\linewidth]{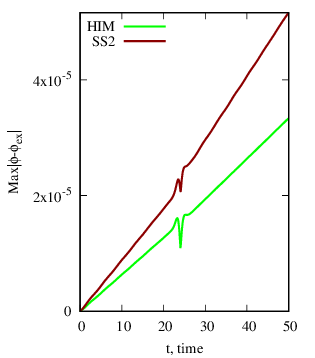}
    \caption{(Collision with stationary soliton on a fully resolved grid) (Left) Error in the solution in $\mathcal{L}_{\infty}$-norm as a function of time step in
    double-logarithmic scale shows second order convergence in $\Delta t$.
    (Right) Absolute error as a function of time, the solitons interact
    at approximately $t = 25$. The error vs time is close to a straight line
    before and after the collision. Its slope, $m$, changes from $m = 6.35\times10^{-7}$ to
    $m = 7.00\times10^{-7}$ for HIM method, and from $m = 8.85\times10^{-7}$ to $m = 1.10\times10^{-6}$ for SS2.}
\label{fig:nabegaet_error}
\end{figure}

\begin{figure}
\includegraphics[width=0.495\linewidth]{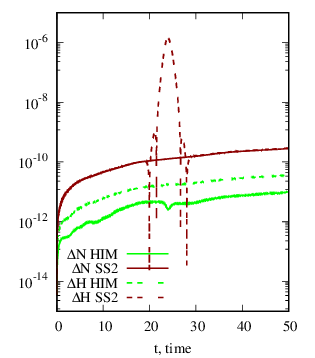}
\includegraphics[width=0.495\linewidth]{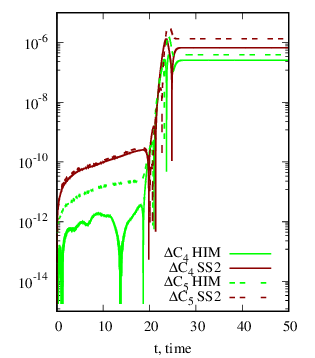}
    \caption{(Collision with stationary soliton on a fully resolved grid) 
    The conserved quantities plotted
    as a function of time over the course of the simulation.
    (Left) The number of particles $\Delta\mathcal{N}$ (solid),
    and the Hamiltonian $\Delta\mathcal{H}$ (dotted), and (right) The integrals $\Delta\mathcal{C}_4$ (solid) and
    $\Delta\mathcal{C}_5$ (dotted) with HIM (green) and SS2 (red).
    SS2 demonstrates a strong peak in error in $\mathcal{H}$ (left panel) at the time of solitons
    interaction. We note that it is a coincidence that lines $\Delta N$ and $\Delta H$ partially overlap each other
    for the SS2 method.
    After the moment of interaction the $\mathcal{C}_4$ and the
    $\mathcal{C}_5$ (right panel) exhibit jump and increase in error with in SS2 and HIM.} 
\label{fig:nabegaet_integrals}
\end{figure}

The initial condition is given by the two-soliton solution formula~\eqref{2soliton} where one of the solitons is moving
towards the other soliton which is at rest. The simulation time is  $T =50$, and over the course of simulation two solitons interact once. We present the
results of the simulation in the Figures~\ref{fig:xtnabegaet} -~\ref{fig:nabegaet_integrals}.

The parameters for this two--soliton solution are given by:
\begin{align}
p_1 = 1.2\,\,\mbox{and}\,\,p_2 = 1.3+i \nonumber \\
a_1 = 2.5+i \,\, \mbox{and} \,\, a_2 = 65+i. \label{icNabegaet}
\end{align}

In the Figure~\ref{fig:xtnabegaet}, the radiation level in SS2 simulation is higher than in the simulation
with HIM method. For both methods we observe that conservation of integrals of motion $\mathcal{H}$,
$\mathcal{N}$, $\mathcal{C}_4$ and $\mathcal{C}_5$ does not imply highly accurate solution in
$\mathcal{L}_{\infty}$-norm as shown in the Figures~\ref{fig:nabegaet_error}--\ref{fig:nabegaet_integrals}.
HIM method gives smaller
$\mathcal{L}_{\infty}$ error in the solution by a factor of $1.5$-$2$ given the same time step size.
In order to compute the $\mathcal{L}_{\infty}$ error we use the exact solution given by the
formula~\eqref{2soliton}.
The simulation time is chosen so that there is a single collision in
the periodic box $[-L,L]$.
The formula~\eqref{2soliton} gives a solution on an infinite line, whereas
the simulation is performed on a periodic box and thus the simulation time must not exceed the time it takes
the solitons to reach the boundary of the box. Moreover, the soliton must still be exponentially
small near the end of the box for the comparison with the exact solution formula to be applicable.

Despite the $\mathcal{L}_{\infty}$ error of the solution not being smaller than $10^{-5}$, we observe
that the integrals of motion $\mathcal{H}$, $\mathcal{N}$ are conserved up to $5\times10^{-10}$. Nevertheless,
at the time of collision we find that
$\Delta \mathcal{H}$ experiences a jump up to $4$ orders of magnitude in SS2 method, while in HIM it is
conserved by construction of the method. Both methods exactly conserve $\mathcal{N}$ aside from accumulation of
round-off errors over the course of simulations. The two nontrivial integrals of motion, $\mathcal{C}_4$ and
$\mathcal{C}_5$ are not conserved exactly, nevertheless we observe that until the time of collision these
quantities vary only in $9$-th decimal place. After the collision these values demonstrate a large jump
(up to four orders of magnitude) in both methods. Unlike the Hamiltonian in SS2 method, these integrals do
not revert to their original values after the collision.

\subsection{Three Solitons Interactions Simulation\label{LongTime}}

It is known that
solitons of the NLSE interact as particles, and interchange momenta during collision~\cite{novikov1984theory}. The details
of the process can be complicated, but once the solitons move sufficiently far from each other, they behave
like separate pulses propagating without change of shape.

In dimensionless units the one--soliton solution is given by~\eqref{1soliton}. For this simulation, the initial condition is the sum of three distinct one--soliton solutions:
\begin{align}
    \Phi(x,t = 0) = \Phi_1+\Phi_2+\Phi_3,
\end{align}
where $\Phi_{1,2,3}$ are given by~\eqref{1soliton} with the following set of parameters:
\begin{align}
    &\lambda_1 = 2.4,\,\,\lambda_2 = 2.9,\,\,\lambda_3 = 3.2, \\
    &v_1 = 0,\,\,v_2 = 0,\,\,v_3 = \frac{2}{3}, \\
    &x_{0,1} = 40,\,\,x_{0,2} = -20,\,\,x_{0,3} = -60,
\end{align}
and zero initial phases. This set of parameters gives us two stationary solitons and one moving. To make sure that we use approximation of a three-soliton solution
on a periodic boundary, we make the overlap between solitons is about $10^{-16}$ and at the boundary $|\Phi(x,t = 0)| \approx 10^{-16}$.

After using the formulae~\eqref{ztaua}, we translate this initial data to dimensional units. In the dimensional units the
characteristic widths, $\tau_c$, and amplitudes, $A$, are given by:
\begin{align*}
    \tau_c &= \frac{1}{\omega_0\sqrt{\lambda}} \approx 50\,\mbox{ps} \\
    A   &= \sqrt{2\lambda}A_0 \approx 2.5\,\mbox{mW}^{1/2}
\end{align*}
and the value of $\lambda$ varies from approximately is $2.4$ to $3.2$. Whereas in the original paper~\cite{PLOpticsLetters2001} the parameters of Gaussian pulses at the end
of the fiber vary in amplitude from approximately $1.0-2.2\,\mbox{mW}^{1/2}$ and have characteristic widths $10-20\,\mbox{ps}$.

The nonlinear time is given by $t_{NL} = \frac{\pi}{|\Phi|^2} = \frac{\pi}{2|\lambda|} \approx 0.5$ which in physical units corresponds to $z_{NL} = t_{NL}l \approx 377$ km.
If transatlantic fiber is considered, this amounts to approximately $26 t_{NL}$. We will illustrate the performance of HIM and SS2, on time scale of $400 t_{NL} \approx 200$
which is still physically relevant.

The solution is computed on a grid of $N = 4096$ points (which corresponds to fully resolved spectrum of solution) with $L = 25\pi$. The fixed point iterations tolerance
is $\varepsilon = 10^{-12}$ for HIM method. The time step for the split step method is chosen to be $\Delta t_{SS2} = \frac{0.8\Delta x ^2}{\pi}$. During simulation time $200$ the solitons interact two times.

\begin{figure}
\includegraphics[width=0.495\textwidth]{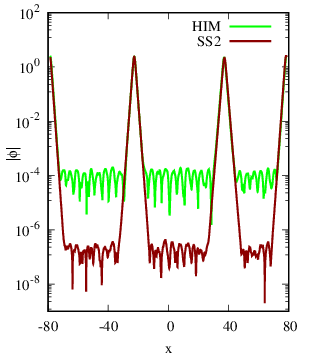}
\includegraphics[width=0.495\textwidth]{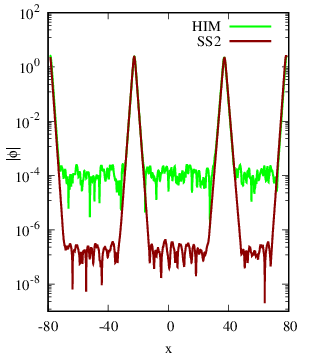}
    \caption{(Left) Absolute value of soliton solution as a function of $x$ at the final
    time of simulation for SS2 (red) with $\Delta t_{SS2} = \frac{0.8\Delta x ^2}{\pi}$ and HIM (green) with $\Delta t_{HIM} = 64 \Delta t_{SS2}$.
    (Right) $|\Phi|$ for SS2 (red) with time step $\Delta t_{SS2} = \frac{0.8\Delta x ^2}{\pi}$ and HIM (green)
    with time step $\Delta t_{HIM} = 128 \Delta t_{SS2}$ at the final time of simulation.
    The green and red lines partially overlap each other because the numerical solutions
    for both methods coincide at solitons peaks.
    The amplitude of radiation in the tails of solitons is about $10^{-7}$ for SS2, and $10^{-4}$ for
    HIM for both time steps. Thess time steps allow HIM to accurately depict the positions of
    the interacting solitons.}
\label{fig:threeSol}
\end{figure}

In this simulation the results are presented in the Figure~\ref{fig:threeSol}, we take $\Delta t_{HIM} = 64 \Delta t_{SS2}$, and due to larger time
step HIM computation time is approximately $5.76$ times smaller. It takes $27.15$ seconds for HIM, and $156.45$ seconds for SS2 to complete the computation on Intel\textregistered Core\texttrademark i7-6700HQ CPU with frequency $2.6$ GHz and $8$ GB RAM in Matlab on a single thread. 

The amplitude of radiation in the tails of solitons is about $10^{-7}$ for SS2, and $10^{-4}$ for
HIM while the time step for HIM is $64$ times larger than for SS2. This time step allows HIM to accurately
depict the positions of the interacting solitons: at the final time the discrepancy in the location of
stationary solitons was less than $\Delta x$. Moreover, if the time step for HIM is increased to
$128\Delta t_{SS2}$ then the discrepancy in the location is still below $2 \Delta x$ and CPU time is $21.30$
seconds on a single thread ($7.35$ times faster than SS2). We note that the amplitude of
radiation in the tails of solitons scales as $\Delta t^2$ for both methods. In exact arithmetic and infinitely
small $\Delta t$ the magnitude of the solution in these regions is exponentially small.

\begin{figure}
\includegraphics[width=0.495\textwidth]{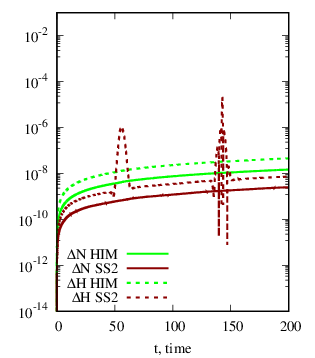}
\includegraphics[width=0.495\textwidth]{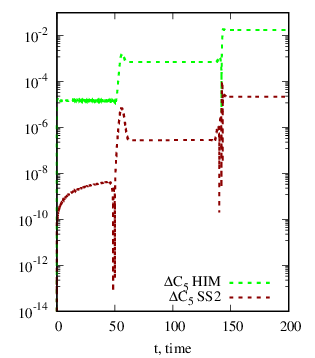}
    \caption{(Left) Error in number of particles, $\Delta N$, and Hamiltonian, $\Delta H$ for SS2 (red) with $\Delta t_{SS2} = \frac{0.8\Delta x ^2}{\pi}$
    and HIM (green) with $\Delta t_{HIM} = 64 \Delta t_{SS2}$. (Right) Error $\Delta C_5$ for SS2 (red) with $\Delta t_{SS2} = \frac{0.8\Delta x ^2}{\pi}$
    and HIM (green) with $\Delta t_{HIM} = 128 \Delta t_{SS2}$.}
\label{fig:threeSolInt}
\end{figure}

In the Figure~\ref{fig:threeSolInt}, we illustrate the conservation of integrals of motion by showing the
difference between the Hamiltonian, the number of particles, and the integrals $C_4$ and $C_5$ at time $t$ and
its value at initial time. We note that the number of particles varies no more than $10^{-7}$ for HIM, and
less than $10^{-8}$ for SS2. The value of Hamiltonian varies no larger than $10^{-7}$ for HIM, however for
SS2 it varies significantly at the time of soliton interaction. We note however, that the accuracy of actual solution
is not representative of these number, and the pointwise error of the numerical solution can be much larger.
The integral $C_4$ is equal to zero in this example, and is not presented in the figure, but the integral $C_5$ is
not zero. It experiences jumps at the time of soliton interactions, and is conserved up to $10^{-2}$
in HIM method due to the much larger time step, $\Delta t_{HIM} = 64 \Delta t_{SS2}$.

\section{Multi-Soliton and Breather Type Solutions}
\subsection{Initial Condition in the Form of $A \operatorname{sech}$}
In this set of simulations, we used the function $A\operatorname{sech}\frac{x}{\sqrt 2}$
with integer $A$ as initial condition. This type of initial conditions was proposed in the
paper of Satsuma and Yajima~\cite{satsuma1974b}. The case $A=1$ corresponds to a stationary
one--soliton solution~\eqref{1soliton}.

We investigate the two soliton solution with $A=2$. For this case, the solution of NLSE has the
form~\cite{satsuma1974b},
\begin{equation}\label{2s}
\Phi(x,t) = 4e^{\frac{-it}{2}}\frac{\cosh\left(\frac{3x}{\sqrt 2}\right)+3e^{-4it}\cosh\left(\frac{x}{\sqrt 2}\right)}{\cosh \left(\frac{4x}{\sqrt 2}\right)+4\cosh \left(\frac{2x}{\sqrt 2}\right)+3\cos(4t)}.
\end{equation}
This solution is periodic in time with period $t_p = 4\pi$.
The equation~\eqref{2s} reduces to $\Phi(x,0) = 2\operatorname{sech(\frac{x}{\sqrt 2})}$ which is used as the initial condition.

The simulation is performed on $N = 2048$ grid points and interval $[-L, L]$ with $L = 12\pi$.
The time of simulation is $T = 40\pi = 10 t_p$, and
time step for both methods is $\Delta t = \frac{0.8}{\pi} \Delta x^2$.
The tolerance for HIM iterations is chosen to be $\epsilon = 10^{-13}$.

\begin{figure}
\includegraphics[width=0.495\textwidth]{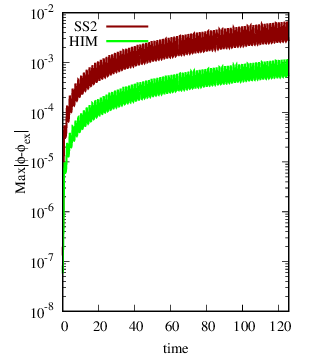}
\includegraphics[width=0.495\textwidth]{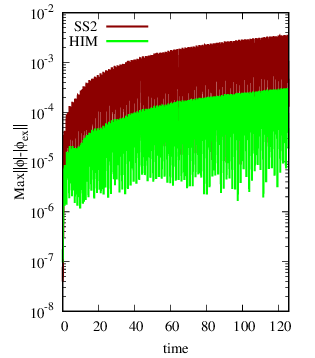}
    \caption{(Simulation of initial condition $2sech\frac{x}{\sqrt 2}$
    on a fully resolved grid)
    (Left) The maximum absolute error of the solution as a function of time.
The HIM method (green) is about $1$ order more accurate than SS2 method (red).
    (Right) $\mathcal{L}_{\infty}$ error of the absolute values of solution.}
\label{fig:2sechsol}
\end{figure}

In the Figure~\ref{fig:2sechsol}, we plot $\mathcal{L}_{\infty}$ norm of the error
in the solution (left panel) and $\mathcal{L}_{\infty}$ error of the absolute value of the solution (right panel).
The $\mathcal{L}_{\infty}$ error in the solutions grows with time for both methods.
The error is smaller by about one order of magnitude in HIM compared to SS2.
The absolute value of the solution is about $2$ orders less accurate in SS2
compared to HIM.

\begin{figure}
\includegraphics[width=0.495\textwidth]{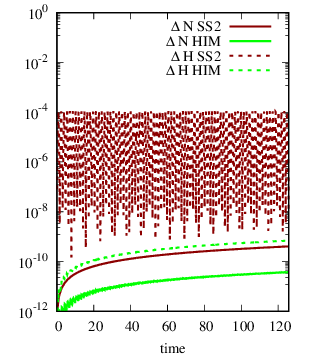}
\includegraphics[width=0.495\textwidth]{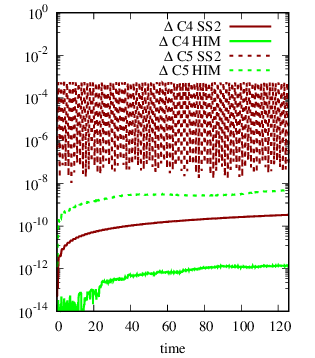}
    \caption{(Simulation of initial condition $2sech\frac{x}{\sqrt 2}$
    on a fully resolved grid)
(Left) The error in integral quantities, $\mathcal{N}$~(solid), and $\mathcal{H}$~(dotted) is about seven orders of magnitude smaller than the error in the solution.
(Right) The error in integral quantities, $\mathcal{C}_4$~(solid),
and $\mathcal{C}_5$~(dotted).}
\label{fig:2sech}
\end{figure}

We show the absolute value of the difference of integrals of motion
at time $t = 0$ and all subsequent times, in the Figure~\ref{fig:2sech}.
Both methods conserve $\mathcal{N}$ equally well.
The Hamiltonian is conserved by HIM up to
$10^{-9}$ and by SS2 up to $10^{-4}$.
There are spikes in $\Delta \mathcal{H}$ from $5 \cdot 10^{-9}$ up to $10^{-4}$ in SS2 method.
The constant of motion $C_4$ is preserved up to $10^{-12}$ by HIM and $10^{-10}$ by SS2.
HIM conserves $\mathcal{C}_5$ up to $10^{-8}$ whereas for SS2 there are spikes
in $\Delta \mathcal{C}_5$ from $10^{-8}$ up to $10^{-4}$.


\subsection{Kuznetsov-Ma Soliton }
Kuznetsov-Ma soliton solution (Kuznetsov~\cite{kuznetsov1977solitons},
Ma~\cite{ma1979perturbed}, Kibler \textit{et al}~\cite{kibler2012observation})
of NLSE has the form,
\begin{equation}\label{ka}
\Phi(x,t) = e^{it}\left[1+\frac{2(1-2a)\cosh(bt)+ib\sinh(bt)}{\sqrt{2a}\cos(wx)-\cosh(bt)}\right]
\end{equation}
where $b = \sqrt{8a(1-2a)}$, $w=2\sqrt{1-2a}$.
It is a periodic function of time with period given by:
\begin{equation}
t_p = \frac{2\pi}{\sqrt{8a(2a-1)}}.
\end{equation}\label{kmperiod}
This formula~\eqref{ka} is taken from Kibler \textit{et al}~\cite{kibler2012observation} and
represents the Kuznetsov-Ma soliton solution for parameter $a > \frac{1}{2}$.

We study the case $a = 1$.
Parameters of the numerical simulation are $N = 1024$ grid points and box size $[-L, L]$ with $L = 12\pi$.
The evolution time is chosen to be $10$ time periods of the solution $T = 7\frac{\pi}{\sqrt 2}$,
and time step is $\Delta t = \frac{0.8}{\pi} \Delta x^2$ for both methods.
The tolerance for HIM iterations is chosen to be $\epsilon = 10^{-13}$.

\begin{figure}
\includegraphics[width=0.495\textwidth]{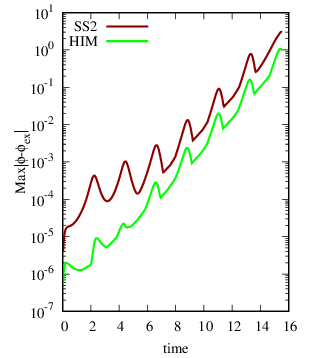}
\includegraphics[width=0.495\textwidth]{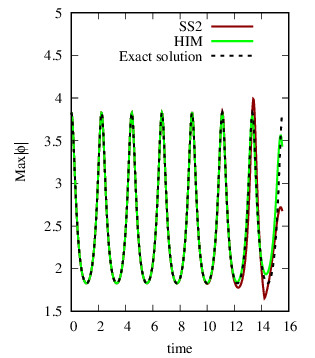}
    \caption{(Kuznetsov-Ma soliton solution on a fully resolved grid)
    SS2 starts to noticeably deviate at about time $12$ and HIM at approximately time $14$
    (Left) The maximum absolute error of the solution as a function of time.
The HIM method (green) is about $1$ order more accurate than SS2 method (red).}
    (Right) Maximum of the absolute values of solution.
    Exact solution (black dotted line) oscillates with period~\eqref{kmperiod}.
\label{KMs}
\end{figure}

In the Figure~\ref{KMs}, we plot the $\mathcal{L}_{\infty}$ error in the solutions (left panel)
and maximum of absolute value of the solution (right panel) as functions of time.
The error in the solution grows with time, and  it is larger in SS2 compared to HIM.
We see that SS2 method loses accuracy earlier (at $t \approx 13$) than HIM ($t \approx 15$).
If we mimic signal propagation in transatlantic fiber, then the final time of computations is
approximately $t\approx 13$, and HIM produces more accurate results at this time.

\begin{figure}
\includegraphics[width=0.495\textwidth]{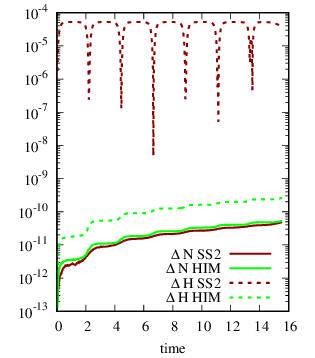}
\includegraphics[width=0.495\textwidth]{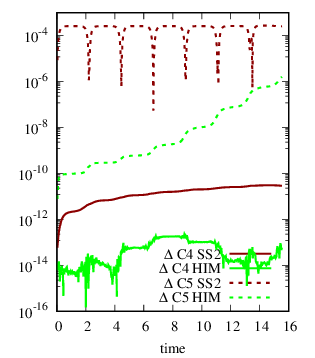}
    \caption{(Kuznetsov-Ma soliton solution on a fully resolved grid)
(Left) The error in integral quantities, $\mathcal{N}$~(solid), and $\mathcal{H}$~(dotted) is about seven orders of magnitude smaller than the error in the solution.
(Right) The error in integral quantities, $\mathcal{C}_4$~(solid),
and $\mathcal{C}_5$~(dotted).}
\label{KMi}
\end{figure}

In the Figure~\ref{KMi}, we show the absolute error in conserved quantities
$\mathcal{N}$, $\mathcal{H}$ (left panel) and $\mathcal{C}_4$, $\mathcal{C}_5$ (right panel).
Both methods conserve the number of particles, $\mathcal{N}$.
The accuracy in the Hamiltonian, $\mathcal{H}$, is about $6$ orders of magnitude different between SS2 and HIM.
Similarly,the difference between SS2 and HIM in $\Delta \mathcal{C}_5$ is about $6$ orders
of magnitude at early times but grows to $2$ orders of magnitude at the end of the simulation.
The integral $\mathcal{C}_4 = 0$ and is conserved by both methods well.
\subsection{Akhmediev Breather}
Akhmediev breather is the solution of NLSE that is periodic in space and
localized in time.
The formula~\eqref{ka} describes Akhmediev breather solution when the parameter $a < \frac{1}{2}$.

We take $a = \frac{1}{4}$ and run simulations on the interval $[-L, L]$
with $L = 2\pi$ and $N = 128$ grid points.
The tolerance for HIM iterations is chosen to be $\epsilon = 10^{-13}$.
Time step for both of methods is $\Delta t = \frac{0.8}{\pi} \Delta x^2$.

\begin{figure}
\includegraphics[width=0.495\textwidth]{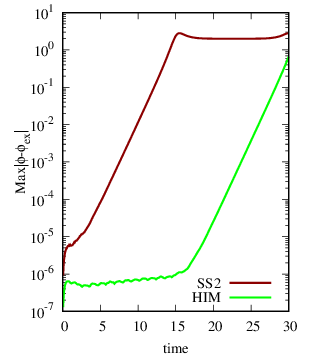}
\includegraphics[width=0.495\textwidth]{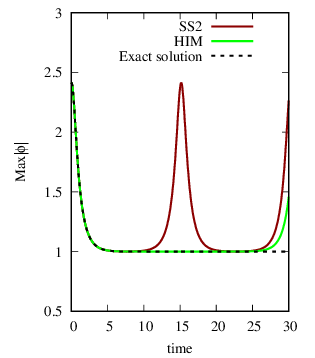}
    \caption{(Akhmediev soliton solution on a fully resolved grid until time $t=30$) (Left) The maximum absolute error of the solution as a function of time.
The error in the SS2 method (red) grows starting from small values of time,
and in HIM (green) stays at about $10^{-6}$ until about time $18$.
(Right) Maximum of absolute value of solution as a function of time.
Exact solution (black dotted line) approaches a constant as time goes to infinity.
SS2 (red) and HIM (green) have oscilations during simulations that deviate
from the exact solution with repietition.}
\label{asz}
\end{figure}

The final time of simulation is $T = 100$, but we show the time interval $t \in [0,30]$
in the Figure~\ref{asz}. The $\mathcal{L}_{\infty}$ error in the solution is on the left panel,
and the maximum of absolute value of the solution is on the right panel.
SS2 fails to produce accurate solution at the time $t = 15$, while
the solution error is approximately $10^{-6}$ for HIM at same instance of time.
The error in the HIM starts to grow from a time $t \approx 17$ and
reaches the error of SS2 at a time $t \approx 30$.
On the right panel, we see that the maximum of absolute value of the exact solution
approaches a constant but the solutions from SS2 and HIM are periodic in time, aka numerical recurrence.
The numerical solutions first approach the constant amplitude solution, but as time increases they diverge
from it.

\begin{figure}
\includegraphics[width=0.495\textwidth]{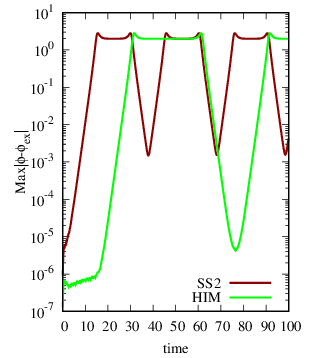}
\includegraphics[width=0.495\textwidth]{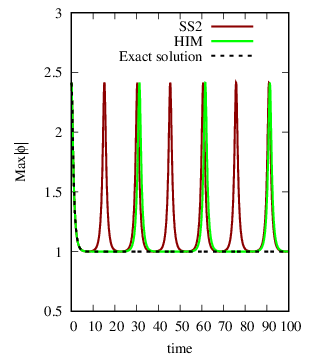}
    \caption{(Akhmediev soliton solution on a fully resolved grid until time $t=100$) (Left) The maximum absolute error of the solution as a function of time.
(Right) Maximum of absolute value of solution as a function of time.
Exact solution (black dotted line) approaches a constant as time goes to infinity.
SS2 (red) and HIM (green) have a couple of oscillations in the solution.
}
\label{as}
\end{figure}

For the simulation time $t = 100$, there are several recurrences in numerical solution
that can be observed in the Figure~\ref{as}.
If we look at the $\mathcal{L}_{\infty}$ error in the solution (left panel),
we see that the error in the solution decreases as solution approaches the exact
solution during the oscillations.
The same behaviour is seen in the maximum of absolute solution as the function of time
(right panel).

\begin{figure}
\includegraphics[width=0.495\textwidth]{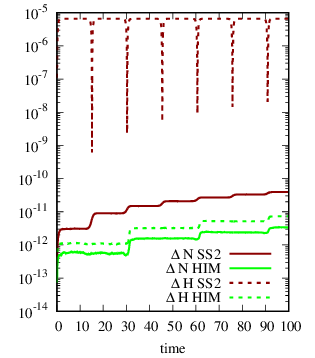}
\includegraphics[width=0.495\textwidth]{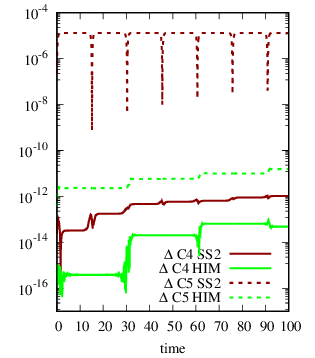}
    \caption{(Akhmediev soliton solution on a fully resolved grid)
(Left) The error in integral quantities, $\Delta \mathcal{N}$~(solid), and $\Delta \mathcal{H}$~(dotted) as a function of time.
The error in $\mathcal{H}$ is about $5$ orders of magnitude is smaller
in HIM compared to SS2, and equivalent in both methods for $\mathcal{N}$.
(Right) The conserved quantities quantities $\Delta \mathcal{C}_4$~(solid),
and $\Delta \mathcal{C}_5$~(dotted) as a function of time.
The error in $\mathcal{C}_5$ is several orders of magnitude smaller in HIM.
The error in $\mathcal{C}_4$ is comparable in both methods.}
\label{ai}
\end{figure}

In the Figure~\ref{ai}, we show the absolute error in constants of motion $\mathcal{N}$, $\mathcal{H}$ (left panel)
and $\mathcal{C}_4$, $\mathcal{C}_5$ (right panel).
SS2 conserves $\mathcal{N}$ and $\mathcal{C}_4$ with good accuracy.
It conserves $\mathcal{H}$ and $\mathcal{C}_5$ up to $10^{-5}$, and
the absolute error in both of these quantities oscillates.
HIM conserves all $4$ constant of motion
$\mathcal{N}$, $\mathcal{H}$, $\mathcal{C}_4$, $\mathcal{C}_5$ up to $10^{-11}$--$10^{-12}$.

\section{\label{s4} Conclusion}
HIM was derived for both NLSE and its generalization MMT model.
We performed detailed comparison of two algorithms for simulation of NLSE: Hamiltonian integration, proposed in~\cite{dyachenko1992optical} and the widely
used split-step method. In all cases Hamiltonian integration demonstrates better conservation of Hamiltonian at the time of soliton collision even for
very large time steps. The other constants of motion $N$, $C_4$ and $C_5$ are conserved better by HIM when the time step is the same or slightly larger than
the one used for split-step method. However, if the time step is increased several orders of magnitude, the accuracy of conservation of integrals of motion in HIM may be lower.
On the other hand, the
pointwise error between the numerical solution and analytic formula is significantly larger than the variation of conserved quantities, which means that integrals of motion reflect
the quality of the solution rather poorly. In experiments we observed this error to be about $10^{-2}$-$10^{-3}$ in the maximum norm. For this reason a
criterion of convergence of fixed point iterations by the number of particles or Hamiltonian~\ref{eq:Hamiltonian}, that was used in the original paper~\cite{dyachenko1992optical}, is suboptimal, and
it is more accurate to control convergence of the residual~\eqref{Residual} as it was proposed in this paper.

However, if the
primary goal is to accurately portray the interaction of solitons over the physically relevant time, such as propagation distance in optical fiber, it is
significantly more advantageous to use the HIM method with large time step rather than SS2 method which requires smaller time step to satisfy the
stability criterion. Violation of stability criterion for SS2 results in complete disintegration of solution for long time simulations~\cite{Lakoba2017}.
In our simulations for $400$ nonlinear times, the time step for HIM is about $64$--$128$ times larger than the instability criterion for SS2. However,
in a simulation for significantly longer time it may lead to accumulation of errors in positioning of the solitons (jitter). For example if one simulates
for $4000$ nonlinear times, the inaccuracy in the soliton position is about $10\Delta x$, and in order to keep the soliton positioning accuracy at $\Delta x$ one would
need to decrease the time step for HIM which results in smaller gains in computation time.

The accurate portrayal of soliton interactions is crucial for the simulation of interactions in soliton gas~\cite{Zakharov1971, Agafontsev2015, turitsyn2009270, turitsyn2010random}, or the fast developing
field of integrable turbulence~\cite{ZakharovEtAl2009}. Both SS2 and HIM approaches are well suited for this. At the same time one should mention that split-step is simpler to
implement and is more efficient memory-wise. The split-step method is explicit, whereas HIM is an implicit method.

As a summary, we would recommend to use Hamiltonian integration method for simulations requiring accurate description of soliton-soliton interactions
or other subtle nonlinear phenomena in Hamiltonian systems especially when computation time is of the essence. Relevance of fast computational algorithms for optical problems can be illustrated by paper~\cite{PLOpticsLetters2001}, \cite{korotkevich2011proof} where massively parallel algorithm for modification of NLSE was proposed and implemented.
For multidimensional turbulence (see for instance~\cite{VladimirovaFalkovich2015}), or for high accuracy short term dynamics, the split-step scheme of the order two, and higher order split
step methods~\cite{Yoshida1990}, \cite{chung2011strong} can be an approach of choice.

\section*{Acknowledgments}
The authors wish to acknowledge gratefully the following contributions:
SAA during her graduate study was partially supported by NSF grants
DMS-1554456 and DMS-1500704; DSA was supported by NSF grant DMS-1716822;
KAO was supported by the Simons Collaboration on Wave Turbulence.  The
work of LPM  was partially supported by the National Science Foundation,
grant DMS-1814619. We thank support of Russian Ministry of Science and
Higher education Grant No. 075-15-2019-1893. Some simulations were
performed  at the Texas Advanced Computing Center using the Extreme
Science and Engineering Discovery Environment (XSEDE), supported by NSF
Grant ACI-1053575.

\appendix
\section{\label{appA} Derivation of the HIM for NLSE}
We consider the equation~\eqref{eq:nlse} where $\gamma = 1$ and with the Hamiltonian~\eqref{eq:Hamiltonian}.
Let $\mathcal{H}^n = \int \left( |\Phi^n_x|^2 - \frac{\gamma}{2}|\Phi^n|^4 \right)$ be the discretized in time Hamiltonian at the $n$-th time step. We consider change of Hamiltonian after one time step $\Delta t$:
\begin{equation}
    \Delta \mathcal{H} = \mathcal{H}^{n+1}-\mathcal{H}^n = I_1 + I_2,
\end{equation} where $I_1 := \int \left( |\Phi^{n+1}_x|^2 -|\Phi^n_x|^2 \right) dx$ and $I_2:=\gamma\int \left( \frac{1}{2}|\Phi^n|^4- \frac{1}{2}|\Phi^{n+1}|^4 \right) dx$.
We consider $I_1$ and $I_2$ separately.

By addition and subtraction to $I_1$ of the following terms, $\frac{1}{2}\Phi^n_x \bar{\Phi}^{n+1}_x$ and
$\frac{1}{2}\Phi^{n+1}_x \bar{\Phi}^{n}_x$, under the integral sign, combining terms and using integration by parts, one gets:
\begin{equation*}
    I_1 = -\frac{1}{2} \int \big( {\bar{\Phi}}^{n+1}_{xx} \Delta\Phi+\Phi^{n}_{xx} \Delta\bar{\Phi}+\Phi^{n+1}_{xx}\Delta\bar{\Phi}
+{\bar{\Phi}}^{n}_{xx} \Delta\Phi \big)dx,
\end{equation*}
here we have introduced $\Delta\Phi = \Phi^{n+1}-\Phi^n$.

By addition and subtraction to $I_2$ of the four following terms, $\frac{\gamma}{2}|\Phi^{n+1}|^2\Phi^n{\bar{\Phi}}^{n+1}$, $\frac{\gamma}{2}|\Phi^{n+1}|^2\Phi^{n+1}{\bar{\Phi}}^n$,
$\frac{\gamma}{2}|\Phi^{n}|^2\Phi^{n+1}{\bar{\Phi}}^n$ and
$\frac{\gamma}{2}|\Phi^{n}|^2\Phi^n{\bar{\Phi}}^{n+1}$, under the integral sign and combining terms, we arrive at
\begin{equation*}
I_2 = -\frac{\gamma}{4}\int \big (\Delta\Phi({\bar{\Phi}}^{n+1}+{\bar{\Phi}}^n)(|\Phi^{n+1}|^2+|\Phi^n|^2)+
\Delta\bar{\Phi}(\Phi^{n+1}+\Phi^n)(|\Phi^{n+1}|^2+|\Phi^n|^2) \big ) dx.
\end{equation*}

After combining the like terms, we arrive at the formula
\begin{multline}\label{H}
    \Delta \mathcal{H} = \frac{1}{2}\int [ \Delta\Phi\left\{-{\bar{\Phi}}^{n+1}_{xx}-{\bar{\Phi}}^n_{xx}-
\frac{\gamma}{2}({\bar{\Phi}}^{n+1}+{\bar{\Phi}}^n)(|\Phi^{n+1}|^2+|\Phi^n|^2)\right\}+\\
\Delta\bar{\Phi}\left\{-\Phi^{n+1}_{xx}-\Phi^n_{xx}-
    \frac{\gamma}{2}(\Phi^{n+1}+\Phi^n)(|\Phi^{n+1}|^2+|\Phi^n|^2)\right\} ] dx.
\end{multline}
If require that the first and second expressions in curly brackets are equal to $\frac{-i\Delta \bar{\Phi}}{\Delta t}$
and $\frac{i\Delta \Phi}{\Delta t}$ respectively, then $\Delta \mathcal{H}$ vanishes.
We note that:
\begin{equation*}
    i\Phi_t = \frac{\delta \mathcal{H}}{\delta \bar{\Phi}},\,\, \text{and} \,\,i\bar{\Phi}_t = -\frac{\delta \mathcal{H}}{\delta \Phi}.
\end{equation*}
We get the following numerical scheme in time:
\begin{equation}
    i\frac{\Phi^{n+1}-\Phi^n}{\Delta t}=-\frac{\left[\Phi^{n+1} + \Phi^n\right]_{xx}}{2}-
\frac{\gamma(\Phi^{n+1}+\Phi^n)(|\Phi^{n+1}|^2+|\Phi^n|^2)}{4}.
\end{equation}

\section{\label{appB} Derivation of the stability condition for NLSE}
In order to solve the equation~\eqref{eq:nlse} one can use the iteration scheme~\eqref{eq:eq4}:
\begin{equation}
    \Phi_k^{n+1,s+1} = \frac{1-\frac{ik^2\Delta t}{2}}{1+\frac{ik^2\Delta t}{2}}\Phi_k^n+\frac{\frac{i\Delta t \gamma}{4}}{1+\frac{ik^2\Delta t}{2}}\hat F\left[(|\Phi^{n+1,s}|^2+|\Phi^n|^2)(\Phi^{n+1,s}+\Phi^n)\right].
\end{equation}
We take $\Phi^{n+1,s+1} = \Phi_0^{n+1}+\delta\Phi^{s+1}$ and $\Phi^{n+1,s} = \Phi_0^{n+1}+\delta\Phi^{s}$ where $\Phi_0^{n+1}$ is the exact solution at the $(n+1)$-st time step.
Let's keep only terms linear in $\delta\Phi^{s+1}$ and neglect terms with small scale perturbations $\delta\Phi^{s}$:
\begin{align}
    \delta\Phi_k^{s+1} &= \frac{\frac{i\Delta t \gamma}{4}}{1+\frac{ik^2\Delta t}{2}}\left[2|\Phi_0^{n+1}|^2+|\Phi^n|^2+\Phi^n\bar{\Phi}_0^{n+1}\right]\delta\Phi_k^s+\nonumber \\
    &\qquad\qquad\qquad\qquad\qquad\qquad+\frac{\frac{i\Delta t \gamma}{4}}{1+\frac{ik^2\Delta t}{2}}\left[(\Phi_0^{n+1})^2+\Phi^n\Phi_0^{n+1}\right]\bar{\delta\Phi_k^s}
\end{align}
Therefore, we can compose the following system of linear equations:
\begin{gather}\label{matrixConv}
\begin{align}
& \begin{bmatrix} \delta \Phi_k^{s+1} \\ \delta \bar{\Phi}_k^{s+1} \end{bmatrix}
    = \nonumber \\
& \begin{pmatrix}
     c\left[2|\Phi_0^{n+1}|^2+|\Phi^n|^2+\Phi^n\bar{\Phi}_0^{n+1}\right] & c\left[(\Phi_0^{n+1})^2+\Phi^n\Phi_0^{n+1}\right] \\
        \bar{c}\left[(\bar{\Phi}_0^{n+1})^2+\bar{\Phi}^n\bar{\Phi}_0^{n+1}\right] & \bar{c}\left[2|\Phi_0^{n+1}|^2+|\Phi^n|^2+\bar{\Phi}^n\Phi_0^{n+1}\right]
    \end{pmatrix}
    \begin{bmatrix}
        \delta \Phi_k^s \\ \delta\bar{\Phi}_k^s
    \end{bmatrix}
\end{align}
\end{gather}
where $c = \frac{\frac{i\gamma\Delta t}{4}}{1+\frac{ik^2\Delta t}{2}}$
We need the matrix on the right hand side of~\eqref{matrixConv} (lets name it $A$) to be a contracting map. As a result, we require its determinant to be smaller than $1$.
From $|\det(A)|< 1$, we can get the condition for iterations convergence of HIM:
\begin{align}
    \Delta t < \dfrac{2}{|\gamma| \sqrt{3} \max(|\Phi^n|^2)}
\end{align}

\section{\label{appC} Interaction of Two-Solitons}
\subsection{Headon Collision of Solitons}\label{HeadOn}
\begin{figure}
\includegraphics[width=2.4in]{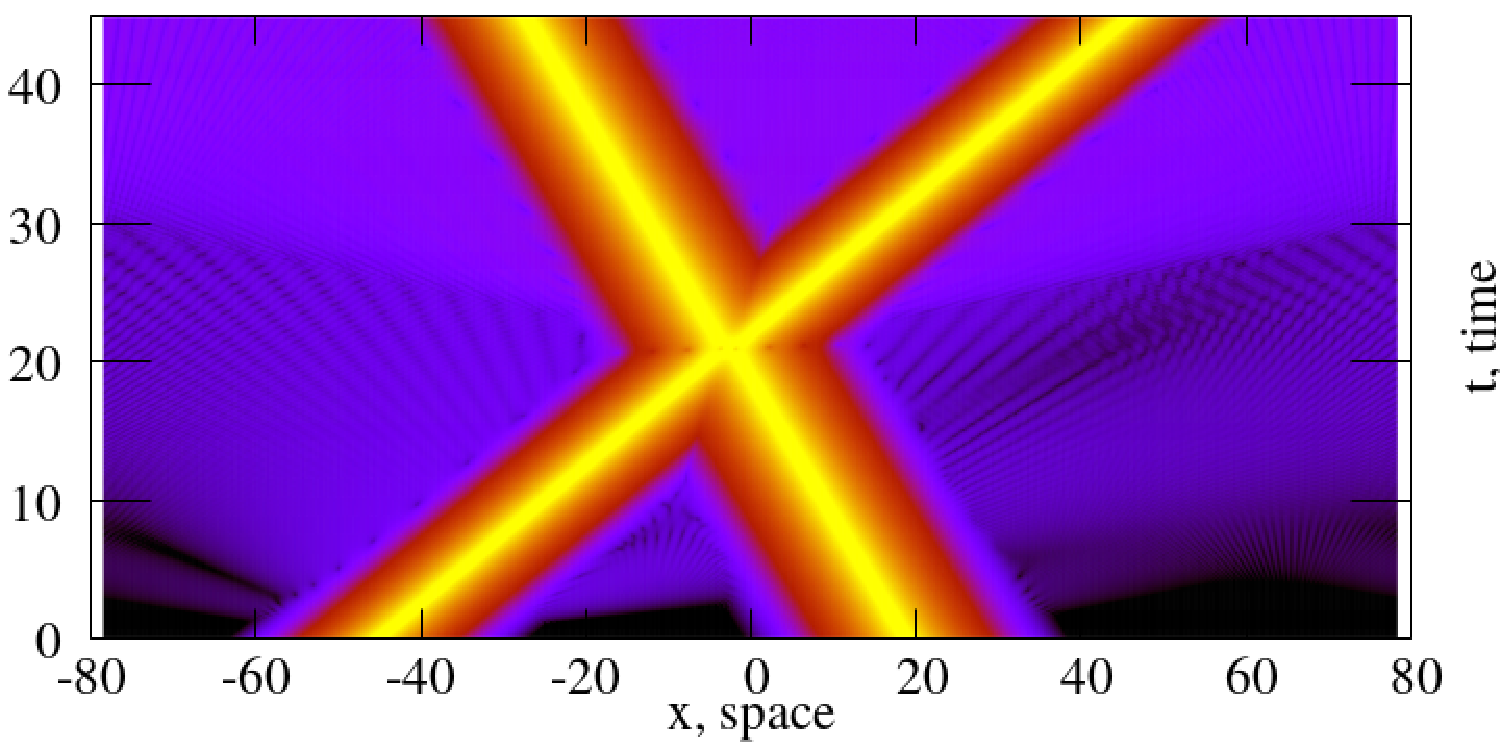}
\includegraphics[width=2.4in]{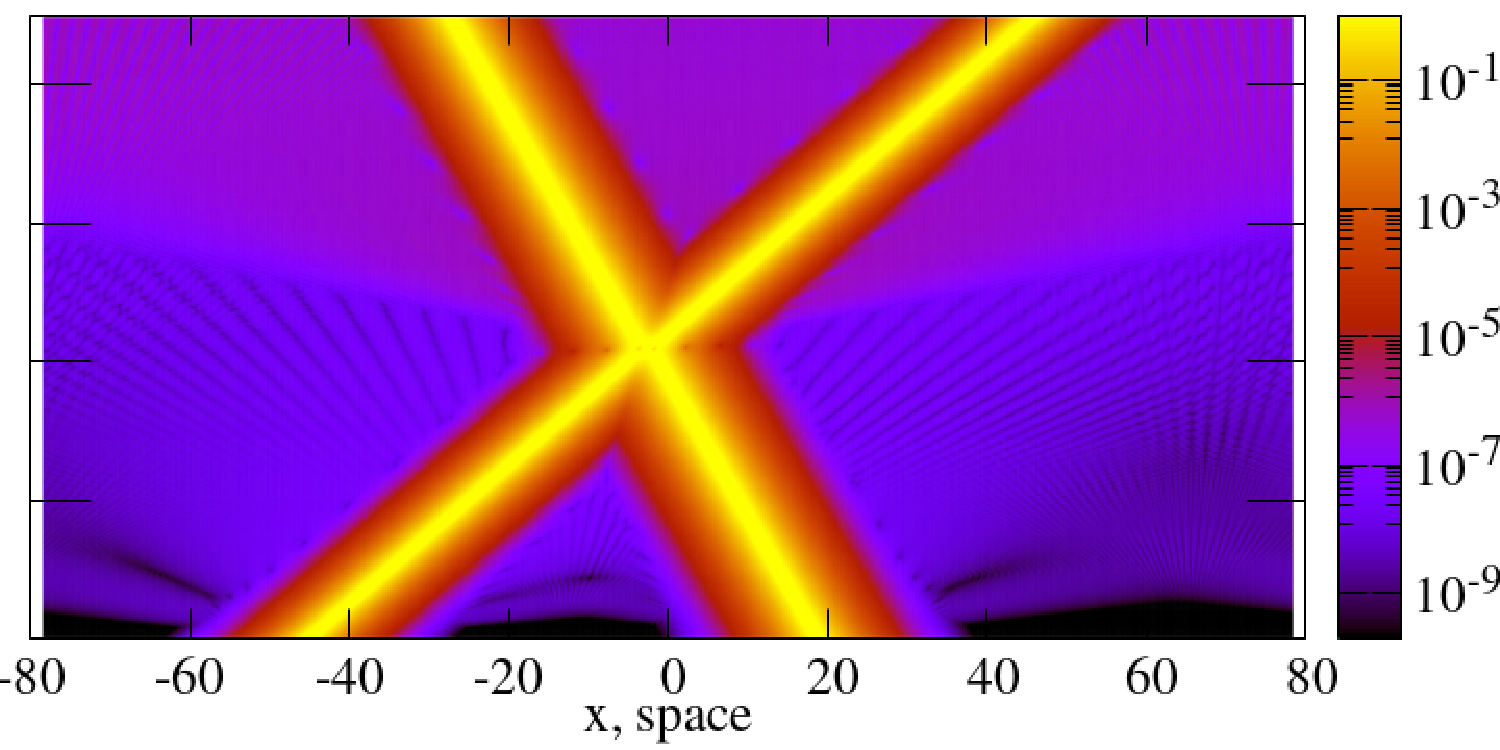}
\includegraphics[width=2.4in]{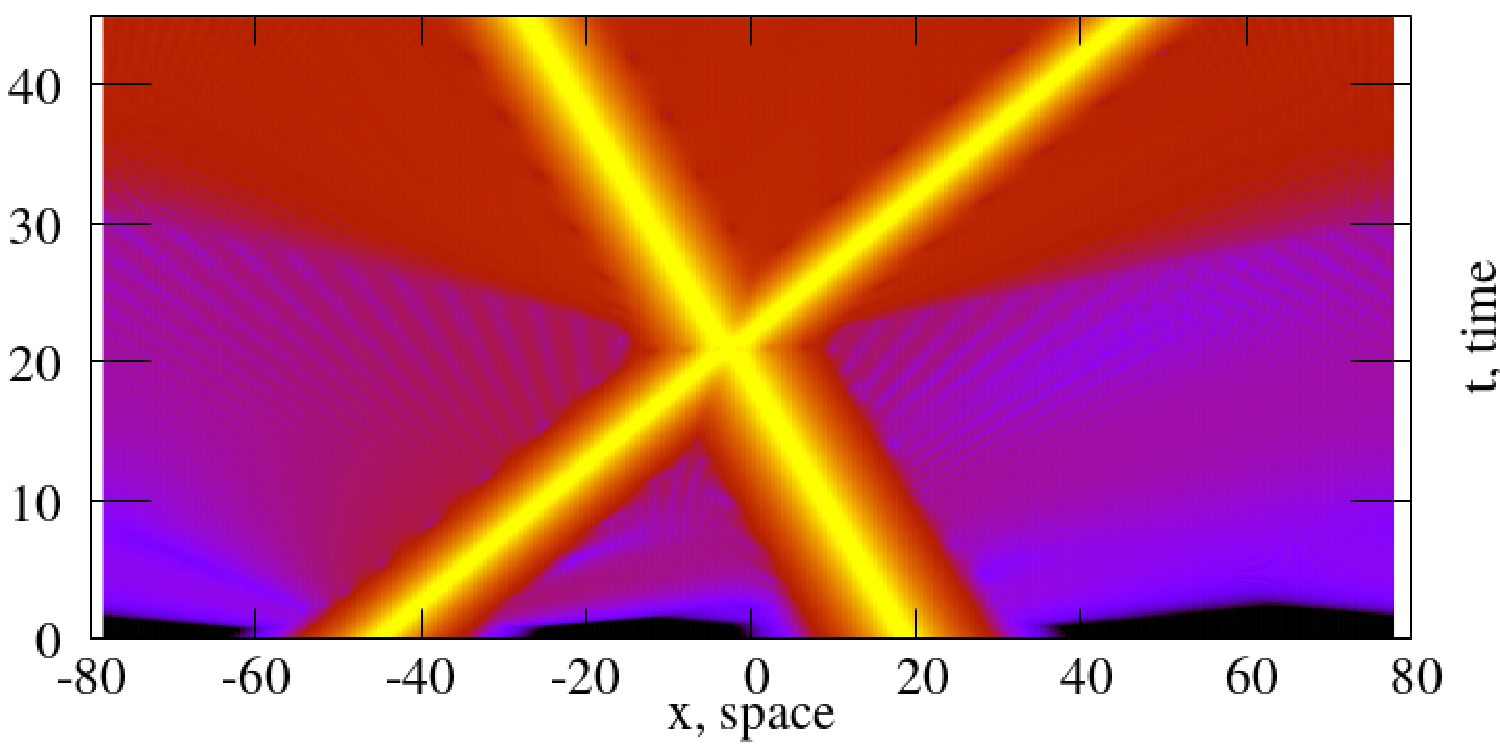}
\includegraphics[width=2.4in]{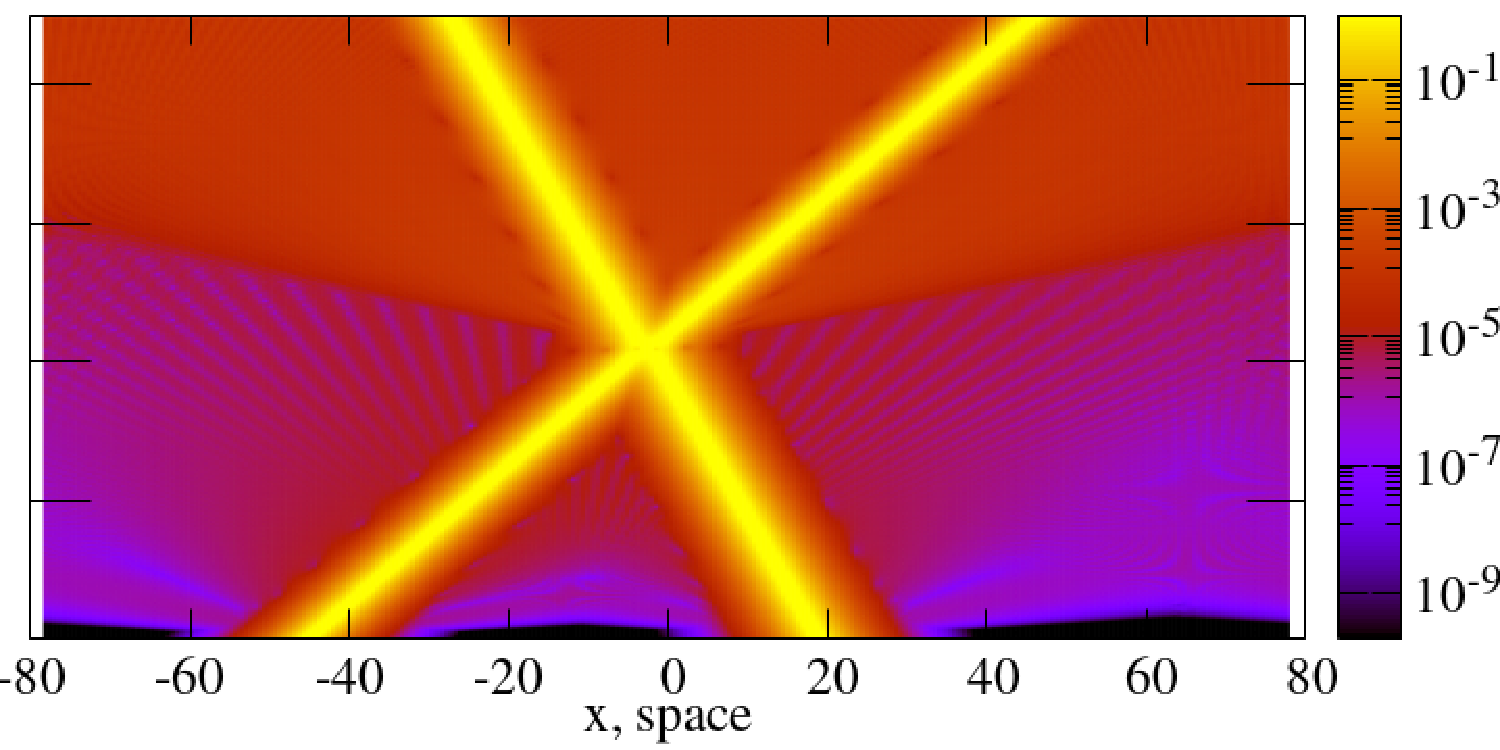}
\caption{(Headon collision of solitons) (Top) Numerical solution for HIM (left) and SS2
    (right) methods on a fully resolved grid $N = 4096$. (Bottom) Numerical solution
    for  HIM (left) and SS2 (right) methods on an underresolved grid with $N = 1024$.}
\label{fig:xtstalkivaet}
\end{figure}

\begin{figure}
    \begin{center}
    \includegraphics[width=1.0\linewidth]{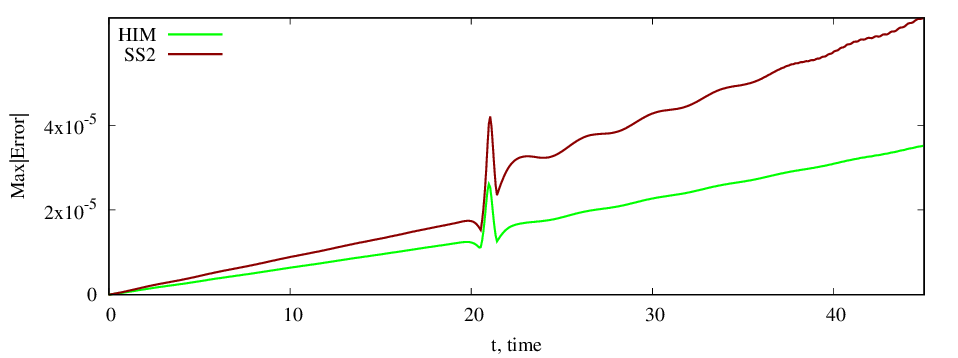}
    \end{center}
    \caption{(Headon collision of solitons on a fully resolved grid) Error in $\mathcal{L}_{\infty}$-norm of the solution vs time computed for
    SS2 (red) and HIM (green) methods. The collision occurs at the time approximately
    $t = 21$ where we observe a spike in the error. The error vs time is close to
    a straight line before and after collision. Its slope, $m$ changes from $m = 8.85\times10^{-7}$
    to $m = 1.5\times10^{-6}$ for SS2 method, and from $m = 6.3\times10^{-7}$ to
    $m = 8.3\times10^{-7}$ for HIM method.
    }
\label{fig:stalkivaet_solErr}
\end{figure}

\begin{figure}
\includegraphics[width=0.495\linewidth]{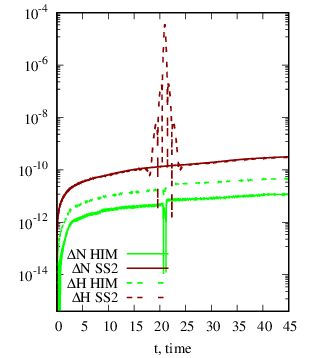}
\includegraphics[width=0.495\linewidth]{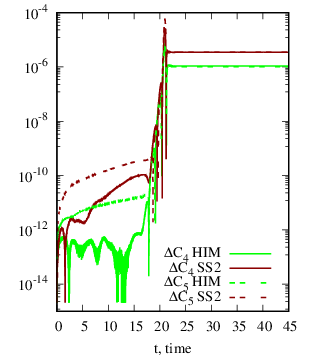}
    \caption{(Headon collision of solitons on a fully resolved grid) The conserved quantities (left) $\Delta \mathcal{N}$ (solid),
    $\Delta \mathcal{H}$ (dotted), and (right) $\Delta \mathcal{C}_4$ (solid), and
    $\Delta \mathcal{C}_5$ (dotted) as a function of time over the course of
    the simulation with HIM (green) and SS2 (red). Note that SS2
    demonstrates a strong peak in error in $\mathcal{H}$ at the time of soliton
    interaction. After the interaction time the $\mathcal{C}_4$, and the
    $\mathcal{C}_5$ exhibit large error with both SS2 and HIM.}
\label{fig:stalkivaet_integrals}
\end{figure}

The initial condition is given by the two-soliton solution formula~\eqref{2soliton} with
solitons moving toward each other. The final time of simulation is $T = 45$,
and two solitons interact once. We present the results of the simulation in
Figures~\ref{fig:xtstalkivaet} -~\ref{fig:stalkivaet_integrals}. The parameters for this
simulation are the following:
\begin{align}
p_1 = 1.2-0.5i\,\,\mbox{and}\,\,p_2 = 1.3+i \nonumber \\
a_1 = -20+i \,\, \mbox{and} \,\, a_2 = 60+i. \label{icHeadOn}
\end{align}

\subsection{Collision with Pursuing Soliton}\label{Pursuing}
\begin{figure}
\includegraphics[width=2.4in]{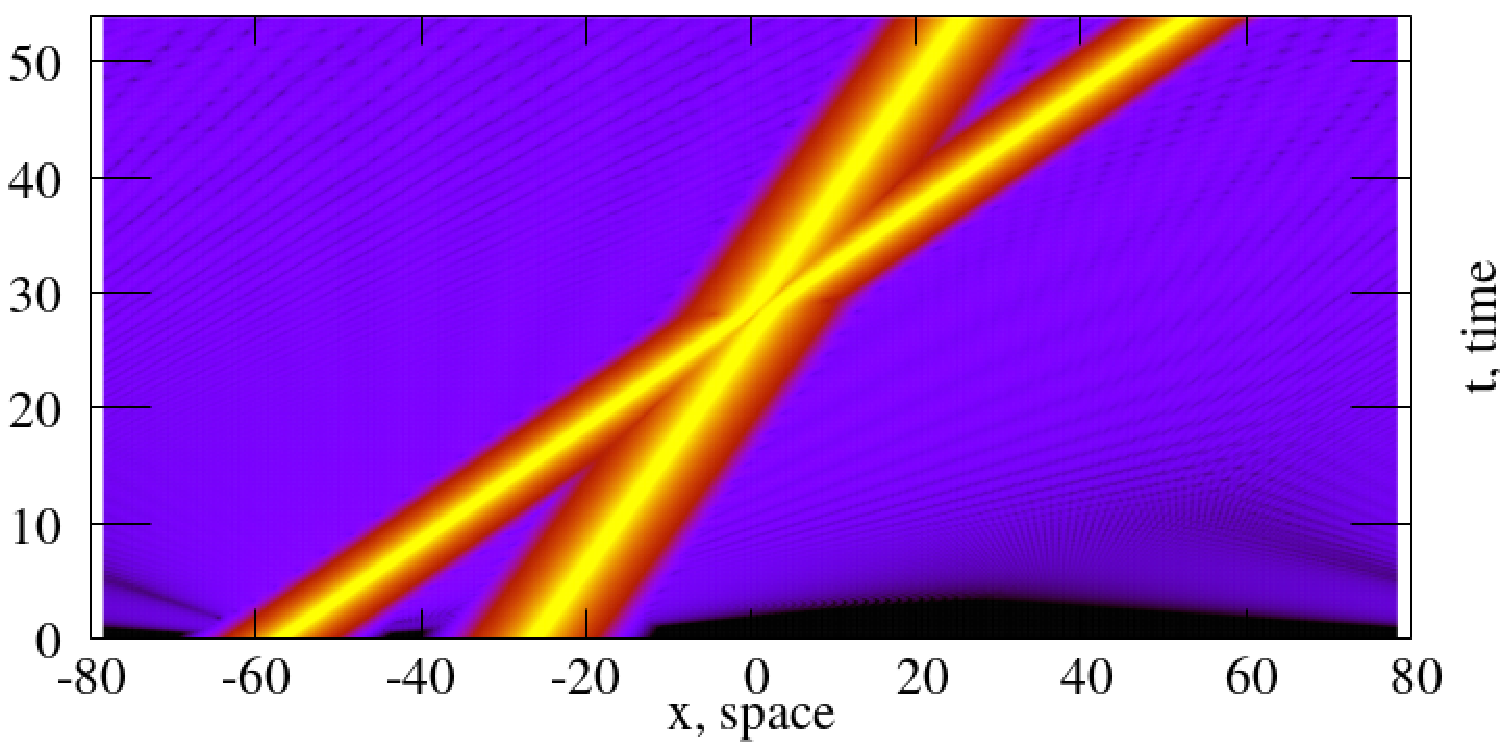}
\includegraphics[width=2.4in]{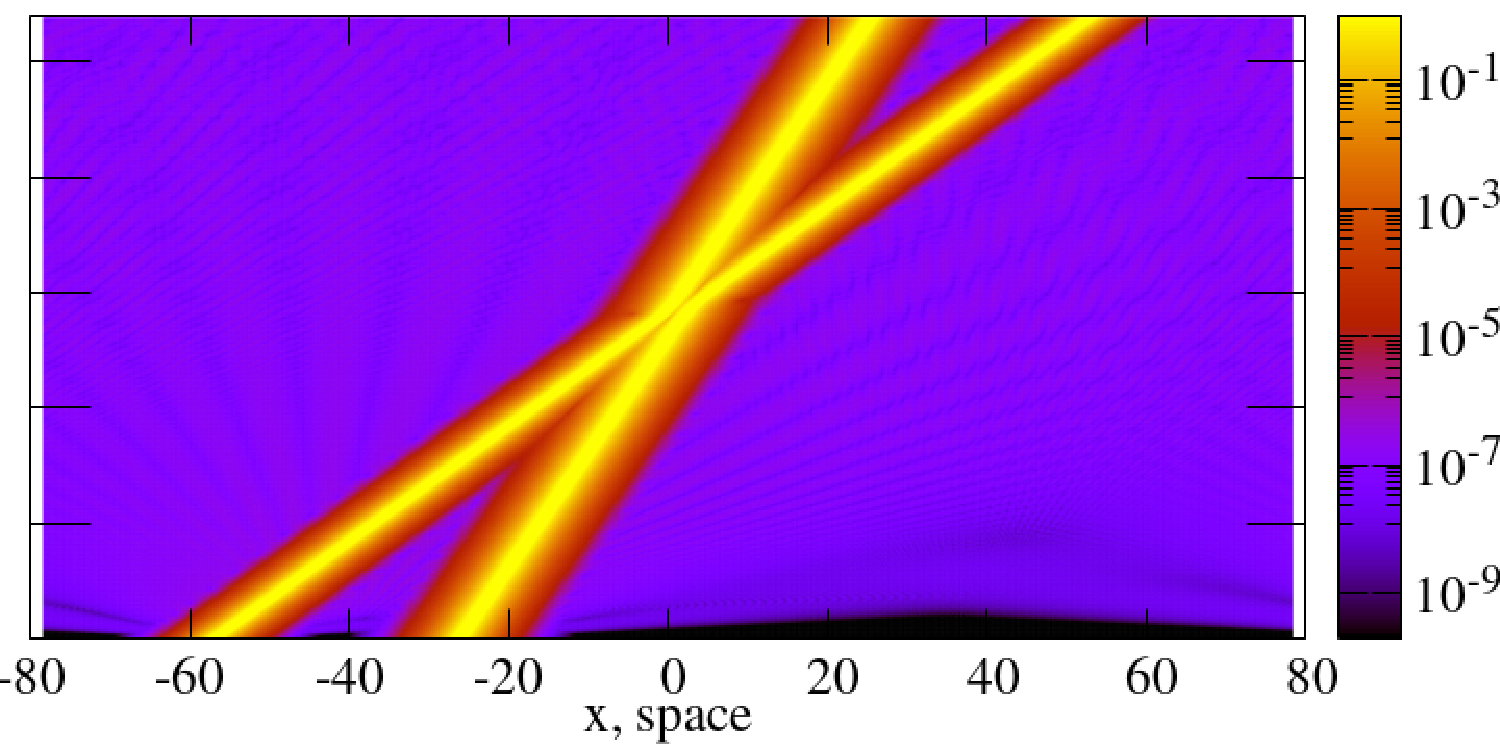}
\includegraphics[width=2.4in]{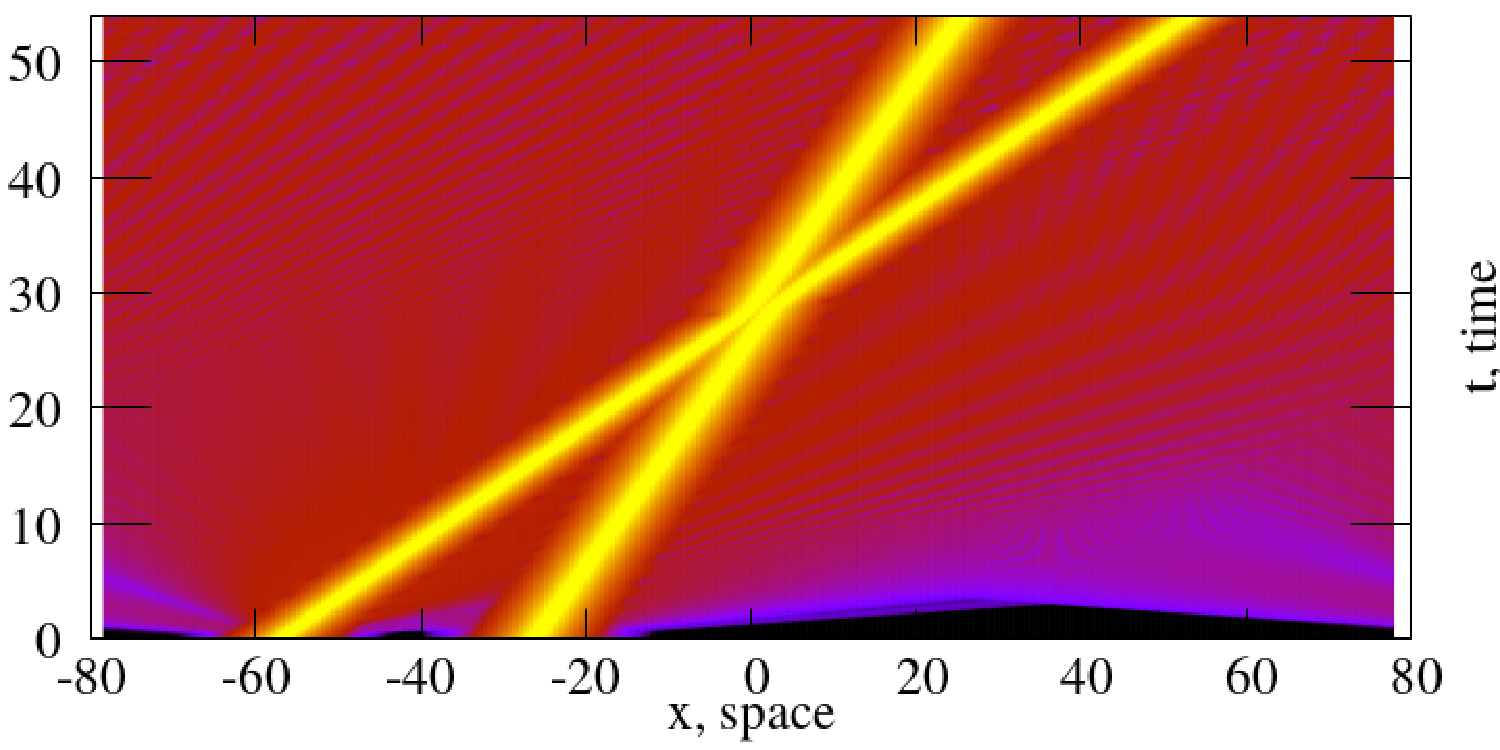}
\includegraphics[width=2.4in]{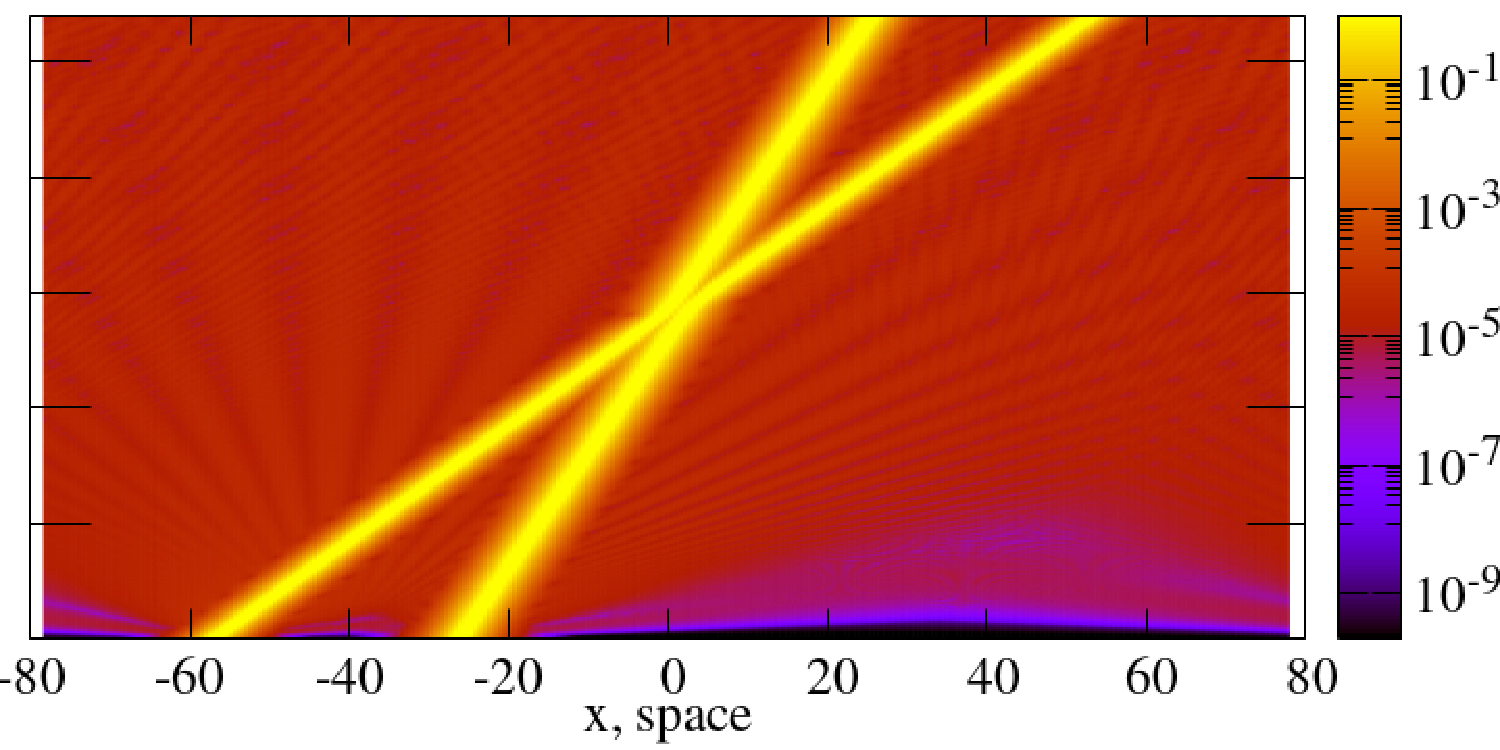}
    \caption{(Collision with pursuing soliton) (Top) Numerical solution for HIM (left) and SS2 (right)
    methods on a fully resolved grid with $N = 4096$ points. (Bottom) Numerical solution for HIM (left)
    and SS2 (right) methods on an underresolved grid with $N = 1024$ points.}
\label{fig:xtdogonyaet}
\end{figure}

\begin{figure}
    \begin{center}
    \includegraphics[width=1.0\linewidth]{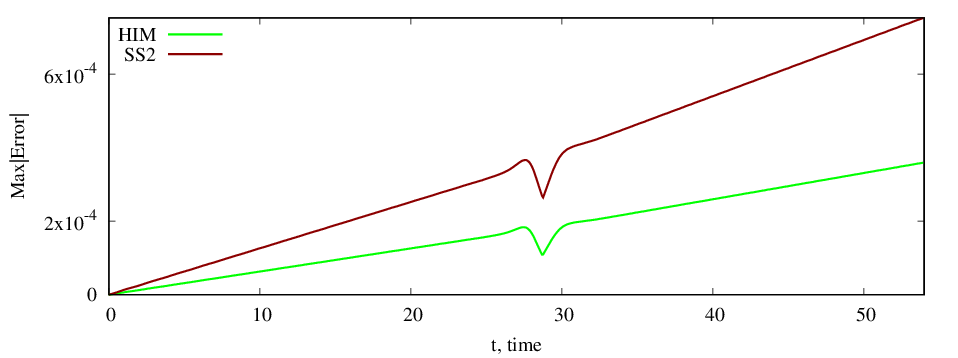}
    \end{center}
    \caption{(Collision with pursuing soliton on a fully resolved grid)  Error in the solution vs time for SS2(red) and HIM(green) methods in the simulation with
    one soliton pursuing the other. The time of collision is approximately $t = 28 $. We observe that
    the slope, $m$ of the straight line of error vs time changes at the collision for both methods. In SS2
    it changes from $m = 1.26\times10^{-5}$ to $m = 1.5\times10^{-6}$, and in HIM the slope changes
    from $m = 6.3\times10^{-6}$ to $m = 7.12\times10^{-6}$.
    }
\label{fig:dogonyaet_solErr}
\end{figure}

\begin{figure}
\includegraphics[width=0.495\linewidth]{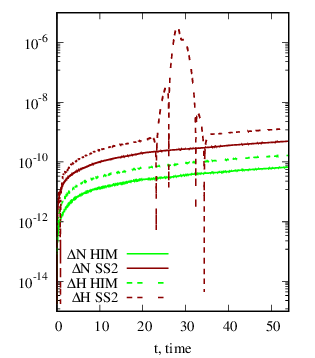}
\includegraphics[width=0.495\linewidth]{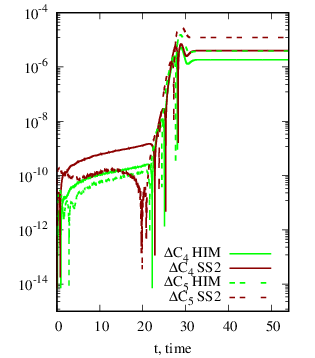}
    \caption{(Collision with pursuing soliton on a fully resolved grid) The error in conserved
    quantities (left) $\Delta \mathcal{N}$ (solid),
    $\Delta \mathcal{H}$ (dotted), and (right) $\Delta \mathcal{C}_4$ (solid), and
    $\Delta \mathcal{C}_5$ (dotted) as a function of time over the course of
    the simulation with HIM (green) and SS2 (red). Note that SS2
    demonstrates a strong peak in error in $\mathcal{H}$ at the time of soliton
    interaction. After the interaction time the $\mathcal{C}_4$, and the
    $\mathcal{C}_5$ exhibit large error with both SS2 and HIM.}
\label{fig:dogonyaet_integrals}
\end{figure}

The initial condition is given by the two-soliton solution formula~\eqref{2soliton} with one soliton pursuing
another soliton. The final time of simulation is $T = 54$. The pursuing soliton
overtakes and interacts with the slower soliton once. The results of this simulation are presented in the
Figures~\ref{fig:xtdogonyaet} -~\ref{fig:dogonyaet_integrals}, and parameters of the initial condition
are as follows:
\begin{align}
p_1 = 1.7+0.5i\,\,\mbox{and}\,\,p_2 = 1.9+i \nonumber \\
a_1 = 50+i \,\, \mbox{and} \,\, a_2 = 110+i. \label{icPursuingSol}
\end{align}

\subsection{Results of Two Soliton Simulations}
In the latter sequence of two simulations involving two-soliton collision, we found
that the radiation level in SS2 simulation has been consistently higher than in simulations
with HIM method. In both methods we observe that conservation of integrals of motion $\mathcal{H}$,
$\mathcal{N}$, $\mathcal{C}_4$ and $\mathcal{C}_5$ does not imply highly accurate solution in
$\mathcal{L}_{\infty}$-norm. In all the cases we found that HIM method gives smaller
$\mathcal{L}_{\infty}$ error in the solution by a factor of at least $1.5$-$2$ with the same time step.
During the simulation time there is a single collision in
the periodic box $[-L,L]$.

Despite the $\mathcal{L}_{\infty}$ error of the solution not being smaller than $10^{-5}$, we observe
that the integrals of motion $\mathcal{H}$, $\mathcal{N}$ are conserved up to $5\times10^{-10}$. Nevertheless,
at the time of collision we found that
$\Delta \mathcal{H}$ experiences a jump up to $5$ orders of magnitude in SS2 method, while in HIM it is
conserved by construction of the method. Both methods exactly conserve $\mathcal{N}$ aside from accumulation of
round-off errors over the course of simulations. The two nontrivial integrals of motion, $\mathcal{C}_4$ and
$\mathcal{C}_5$ are not conserved exactly, nevertheless we observe that until the time of collision these
quantities vary only in $9$-th decimal place. After the collision these values demonstrate a large jump
(up to four orders of magnitude) in both methods. Unlike the Hamiltonian, $\mathcal{H}$, in SS2 method,
these integrals do not revert to their original values after the collision.

\bibliographystyle{elsarticle-num}
\bibliography{bibliography}

\end{document}